\documentclass[paper,notoc]{JHEP3}

\usepackage{epsfig}
\usepackage{graphicx}
\usepackage{feynmp}

\usepackage{amsmath}
\usepackage{amsfonts}
\usepackage{amssymb}
\usepackage{graphicx}

\usepackage{diagrams}

\def\beq{\begin{equation}}
\def\eeq{\end{equation}}
\newcommand{\bea}{\begin{eqnarray}}
\newcommand{\eea}{\end{eqnarray}}

\def\bsp#1\esp{\begin{split}#1\end{split}}

\newcommand{\eps}{\epsilon}

\newcommand{\cI}{\begin{cal}I\end{cal}}

\newcommand{\cA}{{\cal A}}
\newcommand{\cB}{{\cal B}}
\newcommand{\cC}{{\cal C}}
\newcommand{\cS}{{\cal S}}

\newcommand{\cM}{{\cal M}}

\def\cM{{\cal M} }
\def\cN{{\cal N} }
\def\cC{{\cal C} }
\def\cZ{{\cal Z} }
\def\cI{{\cal I} }

\newcommand{\rd}{\mathrm{d}}

\newcommand{\cH}{\begin{cal}H\end{cal}}
\newcommand{\cR}{\begin{cal}R\end{cal}}
\renewcommand{\log}{\ln}

\def\bit#1\eit{\begin{itemize}#1\end{itemize}}
\def\ben#1\een{\begin{enumerate}#1\end{enumerate}}

\newenvironment{sloppyequation}[0]{\sloppy\begin{flushleft}\hspace*{0.75cm}\(}{\)\end{flushleft}\fussy}
\newenvironment{sloppytext}[0]{\sloppy\begin{flushleft}}{\end{flushleft}\fussy}

\newcommand{\beqsloppy}{\begin{sloppyequation}}
\newcommand{\eeqsloppy}{\end{sloppyequation}}
\newcommand{\btxtsloppy}{\begin{sloppytext}}
\newcommand{\etxtsloppy}{\end{sloppytext}}

\newcommand{\li}{\textrm{Li}}
\newcommand{\id}{\textrm{id}}

\title{Hopf algebras, coproducts and symbols:\\
an application to Higgs boson amplitudes}

\author{Claude Duhr\\
Institut f\"ur theoretische Physik, ETH Z\"urich,\\ Wolfgang-Paulistr. 27, CH-8093, Switzerland\\
E-mail:~\email{duhrc@itp.phys.ethz.ch}}

\abstract{We show how the Hopf algebra structure of multiple polylogarithms can be used to simplify complicated expressions for multi-loop amplitudes in perturbative quantum field theory and we argue that, unlike the recently popularized symbol-based approach, the coproduct incorporates information about the $\zeta$ values. We illustrate our approach by rewriting the two-loop helicity amplitudes for a Higgs boson plus three gluons in a simplified and compact form involving only classical polylogarithms.}

\keywords{Feynman integrals, Hopf algebras, polylogarithms, Higgs boson}

\begin{document}

\tableofcontents

\section{Introduction}
\label{sec:intro}
Higher-order quantum corrections to physical observables in perturbative quantum field theory require the evaluation of so-called Feynman integrals, which arise from the integration over the momenta of (unobserved) quanta exchanged in a physical process. For this reason, analytical results for Feynman integrals are not only interesting in their own right but are also of phenomenological relevance in order to make precise predictions for current and future collider experiments.

In many cases Feynman integrals can be expressed in terms of classical polylogarithms and Nielsen polylogarithms~\cite{Nielsen}. In the late nineties it was realized however that for certain multi-loop integrals new classes of transcendental functions arise that can no longer be expressed in terms of the classical polylogarithm functions, \emph{e.g.},~\cite{Remiddi:1999ew,Gehrmann:2000zt,Ablinger:2011te,Vermaseren:2005qc,Moch:2004xu,Vogt:2004mw,Moch:2004pa,Bonciani:2003hc,Bernreuther:2004ih,Bernreuther:2004th,Bernreuther:2005rw,Mastrolia:2003yz,Bonciani:2004qt,Bonciani:2004gi,Czakon:2004wm,Bern:2006ew,Heinrich:2004iq,Smirnov:2001cm,Bork:2010wf,Henn:2010ir,Aglietti:2006tp,Aglietti:2004ki,Aglietti:2004nj,Gehrmann:2001ck,Anastasiou:2006hc,Moch:2002hm,Moch:2002an,Aglietti:2008fe,DelDuca:2009ac, Gehrmann:2001pz,Maitre:2005uu,Maitre:2007kp,Vollinga:2004sn,Davydychev:2000na,Davydychev:2003mv}. It was soon realized that many of the new functions appearing in computations at higher loop orders are in fact special cases of a more general class of functions going under the name of \emph{multiple} polylogarithms in the mathematical literature~\cite{Goncharov:1998, Goncharov:2001}. While multiple polylogarithms do not account for all the classes of functions that can appear in Feynman integral computations (in some cases elliptic functions were encountered, see \emph{e.g.}, Ref.~\cite{Laporta:2004rb}), they are assumed to cover large classes of phenomenologically interesting Feynman integrals.

Just like their classical analogues, not all multiple polylogarithms are independent, but they satisfy (complicated) relations among themselves. These functional equations make it possible to express a given combination of polylogarithms in a multitude of ways with increasing complexity. Thus, while simple and compact analytical results for Feynman integrals are highly desirable, the simplicity of the result might be hidden behind a swath of functional equations. Therefore, a systematic approach and an organizing principle to deal with the functional equations governing the combinatorial structure of (multiple) polylogarithms is a valuable tool to study Feynman integrals at higher loop orders.

In physics a first step in this direction was made in Ref.~\cite{Goncharov:2010jf} with the simplification  of the six-point remainder function in $\cN=4$ Super Yang-Mills computed in Ref.~\cite{DelDuca:2009au, DelDuca:2010zg}. The main tool used to achieve this simplification were the so-called symbols~\cite{Goncharov-Galois}, which provide a way to map the combinatorics of functional equations among iterated integrals to the combinatorics of a certain tensor algebra. Since their introduction, symbols have seen many applications in physics~\cite{Alday:2010ku,Gaiotto:2011dt,Heslop:2011hv,Spradlin:2011wp,Dixon:2011ng,DelDuca:2011ne,DelDuca:2011jm,DelDuca:2011wh,Buehler:2011ev,Brandhuber:2012vm}. In particular, by now the symbols of all two-loop MHV remainder functions~\cite{CaronHuot:2011ky}  and of the two-loop six-point NMHV amplitude~\cite{Dixon:2011nj} are completely known, while at three loops the symbols of the remainder functions for the hexagon in general kinematics~\cite{Dixon:2011pw} and for the octagon in special kinematics~\cite{Heslop:2011hv} are known up to some free parameters that could not be fixed from general considerations. However, only in the latter octagon case an integrated form of the symbol is {also} known. Indeed, while the computation of the symbol of a transcendental function is a straightforward and algorithmic procedure, the inverse problem (sometimes called \emph{integration} of a symbol) of finding in an algorithmic way a function that matches a given symbol is currently still open, despite the fact that advances have been made in how to determine a class of functions that can reproduce a given symbol~\cite{Duhr:2011zq}. One of the problems one encounters during the integration procedure is that the symbol map is non injective. In particular, all terms proportional to (multiple) $\zeta$ value or powers of $\pi$ are mapped to zero by the symbol. 

The aim of this paper is to present an alternative to, or rather an extension of, the naive symbol approach used in physics so far. The cornerstone of our approach is the coproduct on multiple polylogarithms introduced by Goncharov in Ref.~\cite{Goncharov-Galois}, augmented by some ideas  introduced in a recent paper by Brown~\cite{Brown:2011ik}. The coproduct has the advantage that it does not lose as much information as the symbol, while still reproducing the latter in a specific limit. In particular, (multiple) $\zeta$ values are not mapped to zero by the coproduct, thus providing valuable additional information about the function, information that could not be provided by the symbol alone. While we fall short of a full proof of some of our claims, we present accumulating evidence that our approach works in practice by applying it to several non-trivial cases. In particular, we consider the two-loop helicity amplitudes for a Higgs boson plus three gluons computed in Refs.~\cite{Koukoutsakis,Gehrmann:2011aa}, where the results have been expressed as a complicated combination of two-dimensional harmonic polylogarithms. We show that the results of Refs.~\cite{Koukoutsakis,Gehrmann:2011aa} can be rewritten in a compact form that only involves classical polylogarithms with simple rational functions as arguments. This result confirms and extends a similar observation made in Ref.~\cite{Brandhuber:2012vm}, where it was shown that the symbol of the weight-four leading-color contribution of the two-loop helicity amplitudes for a Higgs boson plus three gluons matches the symbol of the form factor of three gluons computed in planar $\cN=4$ Super Yang-Mills.

The outline of the paper is as follows: In Sections~\ref{sec:polylogs} and~\ref{sec:symbols} we provide a short review of our main actors, multiple polylogarithms and symbols. In Section~\ref{sec:algebras} we give a pedestrian introduction to some concepts of modern algebra that are put into action in Section~\ref{sec:Hopf_polylog} where we show how the Hopf algebra of Ref.~\cite{Goncharov-Galois} can be used to simplify complicated expressions involving multiple polylogarithms. This new technique is illustrated on some simple examples in Section~\ref{sec:examples}, while in Section~\ref{sec:Hggg} we apply our tool to rewrite the helicity amplitudes for a Higgs boson plus three gluons computed in Refs.~\cite{Koukoutsakis,Gehrmann:2011aa} in a simplified form. In Section~\ref{sec:conclusion} we draw our conclusions. The appendices contain some technical results omitted throughout the main text.

\section{Multiple polylogarithms}
\label{sec:polylogs}
Analytical results for Feynman integrals are very often expressed in terms of logarithms and (classical) polylogarithms,
\beq\label{eq:classical_polylog}
\ln z = \int_1^z{\rd t\over t} {\rm~~and~~} \li_n(z) = \int_0^z{\rd t\over t}\,\li_{n-1}(t)\,,
\eeq
with $\li_1(z) = -\ln(1-z)$. While these functions are sufficient to describe large classes of Feynman integrals, it is known that especially multi-loop multi-scale integrals can give rise to new classes of functions. 
Among these new classes of functions are the so-called \emph{multiple} polylogarithms, a multi-variable extension of Eq.~\eqref{eq:classical_polylog} defined recursively via the iterated integral~\cite{Goncharov:1998, Goncharov:2001}
 \beq\label{eq:Mult_PolyLog_def}
 G(a_1,\ldots,a_n;z)=\,\int_0^z\,{\rd t\over t-a_1}\,G(a_2,\ldots,a_n;t)\,,\\
\eeq
with $G(z)=1$ and where $a_i, z\in \mathbb{C}$ (they can be either constants or variables in the following). In the special case where all the $a_i$'s are zero, we define, using the obvious vector notation $\vec a_n=(\underbrace{a,\dots,a}_{n})$,
\beq
G(\vec 0_n;z) = {1\over n!}\,\ln^n z\,.
\eeq
The number $n$ of elements $a_i$, counted with multiplicities, is called the {\em weight} of the multiple polylogarithm.

Iterated integrals form a {\em  shuffle algebra}~\cite{Ree:1958}, which allows one to express the product of two multiple polylogarithms of weight $n_1$ and $n_2$ as a linear combination with integer coefficients of multiple polylogarithms of weight $n_1+n_2$,
  \beq\bsp\label{eq:G_shuffle}
  G(a_1,\ldots,a_{n_1};z) \, G(a_{n_1+1},\ldots,a_{n_1+n_2};z) &\,=\sum_{\sigma\in\Sigma(n_1, n_2)}\,G(a_{\sigma(1)},\ldots,a_{\sigma(n_1+n_2)};z),\\
      \esp\eeq
where $\Sigma(n_1,n_2)$ denotes the set of all shuffles of $n_1+n_2$ elements, \emph{i.e.}, the subset of the symmetric group $S_{n_1+n_2}$ defined by
\beq\label{eq:Sigma_def}
\Sigma(n_1,n_2) = \{\sigma\in S_{n_1+n_2} |\, \sigma^{-1}(1)<\ldots<\sigma^{-1}({n_1}) {\rm~~and~~} \sigma^{-1}(n_1+1)<\ldots<\sigma^{-1}(n_1+{n_2})\}\,.
\eeq

Whenever they converge, multiple polylogarithms can equally well be represented as {\em multiple nested sums} (\emph{e.g.}, for $|x_i|<1$)~\cite{Goncharov:1998}
\beq\label{eq:Lim_def}
\textrm{Li}_{m_1,\ldots,m_k}(x_1,\ldots,x_k) = \sum_{n_1<n_2<\dots <n_k} \frac{x_1^{n_1} x_2^{n_2} \cdots x_k^{n_k} }{n_1^{m_1} n_2^{m_2} \cdots n_k^{m_k} } =
 \sum_{n_k=1}^\infty{x_k^{n_k}\over n_k^{m_k}}\,\sum_{n_{k-1}=1}^{n_k-1}\ldots\sum_{n_1=1}^{n_{2}-1}{x_1^{n_1}\over n_1^{m_1}}\,.
\eeq
Note that we are using Goncharov's original summation convention \cite{Goncharov:1998}; other authors define $\textrm{Li}_{m_1,\ldots,m_k}(x_1,\ldots,x_k)$  using the reverse summation convention instead, \emph{i.e.}~$n_1>\dots>n_k$.
The $G$ and Li functions define in fact the same class of functions and are related by
\beq\label{eq:Gm_def}
\textrm{Li}_{m_1,\ldots,m_k}(x_1,\ldots,x_k) = (-1)^k\,G\Big(\underbrace{0,\ldots,0}_{m_k-1},{1\over x_k}, \ldots, \underbrace{0,\ldots,0}_{m_1-1},{1\over x_1\ldots x_k}\Big)\,.
\eeq

It is possible to find closed expressions for special classes of multiple polylogarithms in terms of classical polylogarithm functions, \emph{e.g.}, for $a\neq 0$ we have
\beq\bsp
G(\vec 0_n;z) = {1\over n!}\,\ln^nz, \qquad &G(\vec a_n;z) = {1\over n!}\,\ln^n\left(1-{z\over a}\right),\\
G(\vec 0_{n-1},a;z) = -\textrm{Li}_n\left({z\over a}\right), \qquad & G(\vec 0_{n},\vec a_{p};z) = (-1)^p\,S_{n,p}\left({z\over a}\right)\,,
\esp\eeq
where $S_{n,p}$ denotes the Nielsen polylogarithm~\cite{Nielsen}.

All the notations so far follow the conventions used in physics. Some of the formulas used later on however take a nicer form in a different notation commonly used in the mathematical literature,
\beq\label{eq:math_convention}
I(a_0;a_1,\ldots,a_n;a_{n+1}) = \int_{a_0}^{a_{n+1}}{\rd t\over t-a_{n}}\,I(a_0;a_1,\ldots,a_{n-1};t)\,.
\eeq
The notations~\eqref{eq:Mult_PolyLog_def} and~\eqref{eq:math_convention} are related by (note the reversal of the arguments)
\beq
G(a_n,\ldots,a_1;a_{n+1}) = I(0;a_1,\ldots,a_n;a_{n+1})\,.
\eeq
The iterated integrals defined in Eq.~\eqref{eq:math_convention} are slightly more general than the ones usually defined by physicists, as they allow to freely choose the base point of the integration. It is nevertheless easy to convert every integral with a generic base point $a_0$ into a combination of iterated integrals with base point $0$. An example will clarify this. First, it is easy to see that at weight one we have
\beq
I(a_0;a_1;a_2) = I(0;a_1;a_2) - I(0;a_1;a_0) = G(a_1;a_2) - G(a_1;a_0)  \,.
\eeq
Starting from weight two the relation is more complicated because of the nestedness of the integration, \emph{e.g.},
\beq\bsp
I(a_0&;a_1,a_2;a_3) = \int_{a_0}^{a_3}{\rd t\over t-a_2}\,I(a_0;a_1;t) 
= \int_{a_0}^{a_3}{\rd t\over t-a_2}\,[I(0;a_1;t) - I(0;a_1;a_0)]\\
&\,=I(0;a_1,a_2;a_3) - I(0;a_1,a_2;a_0) - I(0;a_1;a_0)[I(0;a_2;a_3) - I(0;a_2;a_0)]\\
&\,=G(a_2,a_1;a_3) - G(a_2,a_1;a_0) - G(a_1;a_0)[G(a_2;a_3) - G(a_2;a_0)]\,.
\esp\eeq
In the rest of the paper we mostly use the `$I$' notation used in the mathematical literature, as it makes most of the formulas much simpler, keeping in mind that we can easily recover the `$G$' notation via the above procedure.

Just like their classical analogues, multiple polylogarithms satisfy a large number of functional equations among themselves. When expressing a Feynman integral in terms of multiple polylogarithms, we can therefore arrive 
at a complicated combination of multiple polylogarithms, which, if the corresponding functional equations were known, could potentially be reduced to a much simpler expression. While these functional equations are unknown in general, they can often be circumvented in practice by using the so-called symbol, which we will review in the next section.

\section{Symbols}
\label{sec:symbols}
In this section we give a short review of symbols~\cite{Goncharov-simple-Grassmannian}. Symbols were first introduced in physics in Ref.~\cite{Goncharov:2010jf} where they were used to simplify the six-point remainder function in $\cN=4$ Super Yang-Mills computed in Ref.~\cite{DelDuca:2009au, DelDuca:2010zg}. The main idea is to map a (complicated) combination of multiple polylogarithms to a certain tensor algebra over the group of rational functions (the tensors being called symbols) such that, at least conjecturally, all the functional equations among the polylogarithms are mapped to simple algebraic identities in the tensor algebra. 
Currently, two different definitions of symbols are in use in physics,
\begin{enumerate}
\item In Ref.~\cite{Goncharov:2010jf} the symbol of a transcendental function $F_w(x_1,\ldots,x_n)$ of weight $w$ in the variables $x_1,\ldots,x_n$ was defined recursively by considering the total differential of the function $F_w$. More precisely, if the total differential of $F_w$ can be written in the form
\beq\label{eq:diff_eq_generic}
dF_w = \sum_iF_{i,w-1}\,d\ln R_i\,,
\eeq
where $F_{i,w-1}$ are transcendental functions of weight $w-1$ and the $R_i$ are rational functions in the variables $x_1,\ldots,x_n$, then the symbol of $F_w$ can be computed recursively in the weight by
\beq
\cS(F_w) = \sum_i\cS(F_{i,w-1})\otimes R_i\,.
\eeq
Multiple polylogarithms indeed satisfy a differential equation of the type~\eqref{eq:diff_eq_generic}~\cite{Goncharov_Mixed_Tate},
\beq\label{eq:dif_eq_polylog}
dI(a_0;a_1,\ldots,a_n;a_{n+1}) = \sum_{i=1}^nI(a_0;a_1,\ldots,\hat{a}_i,\ldots,a_n;a_{n+1})\,d\ln\left({a_{i+1}-a_i\over a_{i-1}-a_i}\right)\,,
\eeq
where the hat indicates that the corresponding element is omitted. We emphasize though that Eq.~\eqref{eq:dif_eq_polylog} is strictly speaking only valid if all the $a_i$ are generic, \emph{i.e.}, non zero and mutually different. In the non-generic case the differential equation~\eqref{eq:dif_eq_polylog} can take a different form~\cite{Goncharov_Mixed_Tate}.
\item In Ref.~\cite{Duhr:2011zq} an alternative definition of a symbol was given. It was shown that the symbol of a multiple polylogarithm can be obtained by summing over certain dissections of a rooted and decorated polygon associated to a multiple polylogarithm~\cite{GGL:2009}, and the combinatorics of these dissections reproduces precisely the symbol obtained from the recursive procedure of Ref.~\cite{Goncharov:2010jf}.
\end{enumerate}
Both definitions have their advantages and disadvantages. While the recursive definition has the obvious advantage that it is not necessarily restricted to multiple polylogarithms but extends to any class of (transcendental) functions defined through iterated integrals and satisfying a differential equation of the type~\eqref{eq:diff_eq_generic}, the second definition maps the combinatorics of the symbol to the combinatorics of rooted decorated polygons, a correspondence currently only established in the case of polylogarithms. On the other hand, the approach based on polygons is algebraic in nature, and does not make any difference between constants and variables. As an example, the differential equation approach would assign a zero symbol to $\ln 2$ (as $d\ln 2=0$) while $\cS(\ln 2) = 2$ from the polygon approach. As otherwise both definitions are equivalent and give the same answer for multiple polylogarithms, we will in the following not distinguish them any further.

The symbol map $\cS$ fulfills various properties. First, $\cS$ is linear and maps a product of multiple polylogarithms to the shuffle product of their tensors (more precisely, $\cS$ is an algebra homomorphism, see Section~\ref{sec:algebras}). Next, each factor in the symbol is additive with respect to multiplication,
\beq\label{eq:symbol_additivity}
\ldots\otimes(a\cdot b)\otimes\ldots =  \ldots\otimes a\otimes\ldots + \ldots\otimes b\otimes\ldots\,.
\eeq
This implies in particular that
\beq\label{eq:symbol_one}
 \ldots\otimes 1\otimes\ldots = 0\,,
 \eeq
and more generally if $\rho_n$ denotes an $n$-th root of unity,
\beq
 \ldots\otimes \rho_n\otimes\ldots = 0\,.
 \eeq
From Eq.~\eqref{eq:symbol_additivity} and Eq.~\eqref{eq:symbol_one} it is clear that each factor in the tensor product `behaves as a logarithm'. 
 
The first and the last entry of the symbol of a function carry some special information. 
Let us consider a transcendental function $F_w(x_1,\ldots,x_n)$ whose symbol takes the form
\beq
\cS(F_w(x_1,\ldots,x_n)) = \sum_{i_1,\ldots,i_w}c_{i_1,\ldots,i_w}\,\omega_{i_1}\otimes\ldots\otimes\omega_{i_w}\,,
\eeq
where $c_{i_1,\ldots,i_w}$ are (rational) numbers and $\omega_{i_k}$ are rational functions in the $x_i$. The symbol of the derivative of $F_w$ is given by
\beq\label{eq:symbol_derivative}
\cS\left({\partial\over\partial x_k}F_w(x_1,\ldots,x_n)\right) = \sum_{i_1,\ldots,i_w}
c_{i_1,\ldots,i_w}\,\omega_{i_1}\otimes\ldots\otimes\omega_{i_{w-1}}\,\left({\partial\over\partial x_k}\ln \omega_{i_w}\right)\,.
\eeq
In other words, a derivative only acts on the last entry of the symbol.

The first entry of a symbol encodes in a similar way the information about the monodromies (discontinuities) of the function $F_w$. More precisely, if $\cM_{x_k=a}$ is the operator that computes the monodromy of $F_w$ around $x_k=a$, then
 \beq\label{eq:symbol_monodromy}
\cS\left(\cM_{x_k=a}F_w(x_1,\ldots,x_n)\right) = \sum_{i_1,\ldots,i_w}
\left(\cM_{x_k=a}\ln \omega_{i_1}\right)\,
c_{i_1,\ldots,i_w}\,\omega_{i_2}\otimes\ldots\otimes\omega_{i_{w}}\,.
\eeq
 Note that the action of the monodromy operator is trivial on the left-hand side, because it only acts on ordinary logarithms,
 \beq
 \cM_{x_k=a}\ln \omega_{i_1} = \left\{\begin{array}{ll}
 2\pi i\,, & \textrm{if } \omega_{i_1} \textrm{ has a zero for } x_k=a\,,\\
 0\,, & \textrm{otherwise\,.}
 \end{array}\right.
 \eeq
 We prefer nevertheless to write Eq.~\eqref{eq:symbol_monodromy} in this apparently more complicated form in order to exhibit the duality to Eq.~\eqref{eq:symbol_derivative}.
 
So far we have only dealt with the problem of how to compute the symbol of a function.
Indeed, using any of the two definitions we can compute the symbol of any linear combination of products of multiple polylogarithms. Once the symbol has been obtained, 
 the identities~\eqref{eq:symbol_additivity} and~\eqref{eq:symbol_one} allow us to simplify the symbol, which is equivalent to applying functional equations to the original expression. We then have to face the problem, however, of finding a (simpler) function with the same symbol. While there are rules how to compute the symbol of any given combination of polylogarithms, the inverse step of \emph{integrating} the symbol to a function (\emph{i.e.}, of finding a function with the same symbol) is in general much more difficult. In Ref.~\cite{Duhr:2011zq} a prescription was given that allows one to make an educated guess for the class of functions that can give rise to a given symbol. Once such a class of functions has been determined, one can write down a linear combination (with some free coefficients) of these functions and equate their symbols, obtaining in this way a linear system for the coefficients. However, even after this step there is a remaining ambiguity because the symbol map is not injective. As an example, we have
\beq\label{eq:sipi}
\cS(i\pi) = 0 {\rm~~and~~} \cS(\zeta_n) = 0\,.
\eeq
As $\cS$ maps products of functions to shuffle products of tensors, Eq.~\eqref{eq:sipi} implies that all terms proportional to $\zeta$ values and / or $i\pi$ will be mapped to zero by $\cS$. As a consequence, even if we succeed in finding a simpler function with the same symbol as our original function, we are unable to fix the terms proportional to, \emph{e.g.}, $\zeta$ values based on the symbol alone.

The aim of this paper is to introduce a framework similar in spirit to the symbol, but where terms proportional to $\zeta$ values and $i\pi$ are not mapped to zero.
Such a framework should of course retain all the good features of the symbol and reduce to the definition of a symbol in a suitable limit. In the rest of this paper we argue that such a framework is provided by the Hopf algebra of multiple polylogarithms introduced by Goncharov in Ref.~\cite{Goncharov-Galois}, augmented by some ideas inspired by a recent paper by Brown~\cite{Brown:2011ik}.


\section{Algebras, coalgebras and Hopf algebras}
\label{sec:algebras}
The aim of this section is to provide a (pedagogical) review of the algebraic notions used throughout the rest of the paper. The content of this section is standard textbook material in mathematics. We nevertheless include it here because, at least to our knowledge, most of these concepts have only been rarely used in the context of Feynman integral computations. We emphasize that we do not aim at providing a rigorous mathematical exposition of the different topics, but rather content ourselves to provide a pedestrian introduction, and we refer to the dedicated mathematical literature for further details. In particular, we will proceed by analogy with similar mathematical concepts that are of everyday use in physics. Furthermore, we will not be concerned about technical details, such as for example the differences between rings and modules on the one hand, and vector spaces and fields on the other hand. As a consequence, we will use the different notions interchangeably in the following.

\subsection{Algebras: first definition}
We start by reviewing some basic notions about \emph{algebras}.
An algebra over a field $\mathbb{K}$ ($=\mathbb{R}$ or $\mathbb{C}$ in general) is a $\mathbb{K}$-vector space $\cA$ together with a map
\beq\bsp
m: \cA\times \cA &\,\to \cA\\
(a,b) &\,\mapsto m(a,b) \equiv a\cdot b
\esp\eeq
that is associative
\beq
a\cdot (b\cdot c) = (a\cdot b)\cdot c\,,
\eeq
 and has a unit $\varepsilon$,
 \beq\label{eq:unit_element}
 \varepsilon \cdot a = a\cdot \varepsilon = a\,.
 \eeq
Furthermore, the multiplication is compatible with the vector space structure, \emph{i.e.}, distributive,
\beq\label{eq:distributivity}
a\cdot(b+c) = a\cdot b+a\cdot c{\rm~~and~~}(a+b)\cdot c = a\cdot c+b\cdot c\,,
\eeq
and associative with respect to scalars,
\beq\label{eq:scalar_multiplication}
a\cdot(k\, b) = k\,(a\cdot b) {\rm~~and~~} k\,(\ell\,a) = (k\,\ell)\cdot a\,.
\eeq
where $k,\ell\in\mathbb{K}$ are scalars. Let us highlight at this stage some features that will be useful later on. First, the distributivity~\eqref{eq:distributivity} implies that the multiplication $m$ is in fact a bilinear map from $\cA\times\cA$ to $\cA$. Second, as a consequence of the compatibility with the scalar multiplication, Eq.~\eqref{eq:scalar_multiplication}, we will assume from now on that the field of scalars $\mathbb{K}$ is part of the algebra itself, \emph{i.e.}, $\mathbb{K}$ can be embedded into $\cA$. Under this assumption it is easy to see that we can identify the unit element $\varepsilon$ in Eq.~\eqref{eq:unit_element} with the unit element $1\in\mathbb{K}$.

While this definition of an algebra is presumably familiar to most physicists, it is not well suited to understand the link to coalgebras and Hopf algebras. For this reason, we will now reformulate the above definition in terms of tensor products of vector spaces. 

\subsection{Tensor products of vector spaces}
As the concept of tensor product used in the following is different from the definition commonly used in  physics, let us review the mathematical definition of a tensor product.

Consider three vector spaces $U$, $V$ and $W$. A standard textbook result then states that there is a unique vector space, called the tensor product $U\otimes V$ of $U$ and $V$, and a unique \emph{bilinear} map $\tau: U\times V\to U\otimes V$ such that for every \emph{bilinear} map $\beta:U\times V\to W$ there is a unique \emph{linear} map $\mu$ such that
\beq\label{eq:def_tensor_product}
\beta = \mu\,\tau\,,
\eeq
where $\mu\,\tau$ denotes the composition of $\mu$ and $\tau$.
The bilinear map $\tau$ simply assigns to two vectors $a$ and $b$ their tensor product, \emph{i.e.}, $\tau(a,b) =a\otimes b$. In other words, we can reformulate Eq.~\eqref{eq:def_tensor_product} in an equivalent, but more accessible, form: for every \emph{bilinear} map $\beta$ there is a unique \emph{linear} map $\mu$ such that
\beq\label{eq:muab}
\beta(a,b) = \mu(a\otimes b)\,.
\eeq
According to Eq.~\eqref{eq:muab} it is always possible to break the bilinearity of $\beta$ up into the bilinearity of the tensor product and the linearity of $\mu$. As an example, we have
\beq\bsp
\beta(k\,a+\ell\,b,c) &\,= \mu((k\,a+\ell\,b)\otimes c) = \mu(k\,(a\otimes c) + \ell\,(b\otimes c)) \\
&\,= 
k\,\mu(a\otimes c)+\ell\,\mu(b\otimes c) = k\,\beta(a,c) + \ell\,\beta(b,c)\,.
\esp\eeq
Before applying this result to the definition of algebras,
let us take the opportunity to introduce a diagrammatic tool useful to describe algebraic structures. If we represent each map between vector spaces by an arrow, \emph{e.g.}, 
\beq
U\times V \stackrel{\tau}{\longrightarrow} U\otimes V\,, \qquad U\times V \stackrel{\beta}{\longrightarrow} W\,, \qquad U\otimes V\stackrel{\mu}{\longrightarrow} W\,,
\eeq
then the relation~\eqref{eq:def_tensor_product} between $\tau$, $\beta$ and $\mu$ is conveniently described by the following \emph{commutative diagram},
\beq
\begin{diagram}
U\times V& & \rTo^{\tau} && U\otimes V \\
 & \rdTo(4,2)_{\beta} &&& \dTo_{\mu} \\
 &&&& W
\end{diagram}
\eeq
The word `commutative' refers in this context to the fact that we can take any path through the diagram from  $U\times V$ to $W$ and we always arrive at the same result. This is precisely the content of Eq.~\eqref{eq:def_tensor_product}.

\subsection{Algebras: second definition}
We have seen that the distributivity of the multiplication $m$ implies that $m$ is in fact a bilinear map from $\cA\times\cA$ to $\cA$. Thus, following our considerations on tensor products, there is a unique linear map $\mu:\cA\otimes\cA \to \cA$ such that
\beq
a\cdot b = m(a,b) = \mu(a\otimes b)\,.
\eeq
The associativity of the tensor product can then be summarized by the following property of the linear map $\mu$,
\beq\label{eq:associativity}
\mu(\id\otimes\mu) = \mu(\mu\otimes\id)\,.
\eeq
Indeed, acting with $\mu(\id\otimes\mu)$ on $a\otimes b\otimes c$ gives\footnote{For two functions $f$ and $g$ we define $(f\otimes g)(a\otimes b) = f(a)\otimes g(b)$.}
\beq
\mu(\id\otimes\mu)(a\otimes b\otimes c) = \mu(a\otimes \mu(b\otimes c)) = \mu(a\otimes (b\cdot c))
=a\cdot(b\cdot c)\,,
\eeq
while acting with $\mu(\mu\otimes\id)$ gives
\beq
\mu(\mu\otimes\id)(a\otimes b\otimes c) = \mu(\mu(a\otimes b)\otimes c) = \mu((a\cdot b)\otimes c)
=(a\cdot b)\cdot c\,.
\eeq
In order to make the transition to coalgebras easier, it is useful to rephrase Eq.~\eqref{eq:associativity} in terms of commutative diagrams. It is easy to check that Eq.~\eqref{eq:associativity} is equivalent to the commutativity of the diagram
\beq
\begin{diagram}\label{eq:associativity_diagram}
\cA\otimes \cA\otimes \cA && \rTo^{\id\otimes \mu} && \cA\otimes \cA \\
\dTo^{\mu\otimes \id} &&&&\dTo_{\mu}\\
\cA\otimes \cA && \rTo^\mu &&\cA
\end{diagram}
\eeq
The existence of a unit element can also be recast into this new language. Indeed, we previously assumed that the field of scalars is embedded into the algebra. This implies the existence of a map $\epsilon:\mathbb{K} \to\cA$ that assigns to each scalar $k$ and element $\epsilon(k)\in\cA$. Compatibility with the scalar multiplication forces us to require that $\epsilon$ be linear and that, for any vector $a$, multiplication with the scalar $k$ or the vector $\epsilon(k)$ gives the same result,
\beq\label{eq:unit_scalar}
\epsilon(k)\cdot a = k\, a {\rm~~and~~} \epsilon(k)\cdot\epsilon(\ell) = \epsilon(k\,\ell)\,.
\eeq
In terms of commutative diagrams Eq.~\eqref{eq:unit_scalar} is equivalently expressed as
\beq
\begin{diagram}\label{eq:unit_diagram}
\mathbb{K}\otimes \cA \\
\dTo^{\epsilon\otimes \id} & \rdTo(4,2)^{s} \\
\cA\otimes \cA &&\rTo^{\mu} && \cA
\end{diagram}
\eeq
where $s$ denotes the scalar multiplication of a vector and a scalar, $s(k,a) = k\,a$. Note that the embedding $\epsilon$, together with Eq.~\eqref{eq:unit_scalar}, implies the existence of the unit $\varepsilon \equiv \epsilon(1)$ in the usual sense, because
\beq
\epsilon(1) \cdot a = 1\, a = a\,.
\eeq
For this reason, the map $\epsilon$ is usually referred to as the \emph{unit} of the algebra.

Another recurrent theme in the study of algebraic structures is the study of the structure-preserving maps (the so-called \emph{homomorphisms}). If $\cA$ and $\cB$ are algebras, then a homomorphism from $\cA$ to $\cB$ is a linear map $\phi$ that preserves the multiplication,
\beq
\phi(a\cdot b) = \phi(a)\cdot\phi(b)\,.
\eeq

At this stage we can identify the multiple polylogarithms as an algebra: the multiplication is given by the shuffle product, while the scalars are given by (rational) numbers. In addition, the shuffle product  preserves the weight (the product of two polylogarithms of weight $n_1$ and $n_2$ gives a linear combination of polylogarithms of weight $n_1+n_2$). This feature is formalized 
by the notion of a \emph{graded} algebra, \emph{i.e.}, an algebra that is
a direct sum as a vector space,
\beq
\cA = \bigoplus_{n=0}^\infty\cA_n\,,
\eeq
such that the multiplication preserves the grading,
\beq
\cA_m\cdot \cA_n \subset \cA_{m+n}\,.
\eeq
An element of $\cA_n$ is said to be of \emph{weight} or \emph{grade} $n$. The multiple polylogarithm algebra is thus graded by the weight. In the following we always assume that the weight 0 part of a graded algebra coincides with the field of scalars, $\cA_0=\mathbb{K}$. Note that this is in agreement with our naive expectation in the case of multiple polylogarithms: the weight 0 part of the algebra consists of all objects of transcendental weight 0, \emph{e.g.}, rational numbers or rational functions.

To summarize, we arrive at the following conclusion: an algebra $\cA$ is a vector space equipped with a linear map $\mu:\cA\otimes\cA\to\cA$ and an algebra homomorphism $\epsilon:\mathbb{K}\to\cA$ such that $\mu$ satisfies the associativity condition expressed through Eq.~\eqref{eq:associativity} or, equivalently, through the commutative diagram~\eqref{eq:associativity_diagram}. The multiple polylogarithms, equipped with the shuffle product, then form an algebra graded by the weight. 

So far we have only formalized the algebraic structure of multiple polylogarithms in an unusual way. In Section~\ref{sec:Hopf_polylog} we will see that multiple polylogarithms carry another structure, a so-called Hopf algebra structure, which is the topic of the rest of this section.

\subsection{Coalgebras as the duals of algebras}
In this section we introduce coalgebras and coproducts, one of the main topics in the rest of this paper. We refrain from giving a mathematically precise axiomatic definition of a coalgebra, and we rather proceed by analogy with similar concepts of everyday use in physics: vector spaces and hermitian conjugates of linear maps.

Let us start by reviewing some textbook material on the duals of vector spaces.
Consider two vector spaces $V$ and $W$ and denote their duals\footnote{In quantum mechanics, the elements of $V$ are called `ket's, while the elements of $V^\ast$ are called `bra's}, \emph{i.e.}, the vector spaces of all linear forms on $V$ and $W$, by $V^\ast$ and $W^\ast$. Let $A$ be a linear map from $V$ to $W$. It is well-known that $A$ induces a linear map between the duals, the hermitian conjugate $A^\dagger$, in the opposite direction,
\beq\label{eq:dualization}
V \stackrel{A}{\longrightarrow} W {\rm~~and~~} W^\ast \stackrel{A^\dagger}{\longrightarrow} V^\ast\,.
\eeq
Eq.~\eqref{eq:dualization} provides us with a diagrammatic rule to derive the commutative diagram describing some algebraic structure when replacing all the vector spaces by their duals. Indeed, when dualizing, we 
\begin{enumerate}
\item replace each vector space by its dual,
\item replace each linear map by its dual (= hermitian conjugate),
\item reverse all the arrows in the diagram.
\end{enumerate}
If we upgrade $V$ to an algebra, then it comes equipped with two linear maps, the multiplication $\mu$ and the unit $\eps$, so we can consider their `hermitian conjugates', \emph{i.e.}, the structures induced on the dual space $V^\ast$ by the multiplication and the unit.

A \emph{coalgebra} is defined as the dual of an algebra, \emph{i.e.}, if $\cA$ is an algebra with multiplication $\mu: \cA \otimes \cA \to \cA$ and unit $\epsilon:\mathbb{K}\to \cA$, its dual $\cC=\cA^\ast$ is equipped with linear maps $\Delta = \mu^\dagger:\cC \to \cC\otimes \cC$ (the \emph{comultiplication}) and $\eta = \eps^\dagger : \cC\to \mathbb{K}$ (the \emph{counit}). The properties of the algebra operations (associativity and unit) will also be reflected in the coalgebra $\cC$. Using the diagrammatic rule given in Eq.~\eqref{eq:dualization}, we can easily obtain the commutative diagrams that describe the duals to the associativity~\eqref{eq:associativity_diagram} and the unit~\eqref{eq:unit_diagram} (the \emph{coassociativity} and the \emph{counit}),
\beq\bsp
\begin{diagram}
\cC\otimes \cC\otimes \cC && \lTo^{\id\otimes \Delta} && \cC\otimes \cC \\
\uTo^{\Delta\otimes \id} &&&&\uTo_{\Delta}\\
\cC\otimes \cC && \lTo^\Delta &&\cC
\end{diagram} & \qquad\qquad
\begin{diagram}
\mathbb{K}\otimes \cC \\
\uTo^{\eta\otimes \id} & \luTo(4,2)^{s^\dagger} \\
\cC\otimes \cC &&\lTo^{\Delta} && \cC
\end{diagram} 
\esp\eeq
The coassociativity can also be written in equations as
\beq\label{eq:coassociativity}
(\id \otimes\Delta)\Delta = (\Delta \otimes\id)\Delta\,.
\eeq
As the coassociativity will be extensively used in the remainder of this paper, let us elaborate on it for a while. While the multiplication $\mu$ corresponds to the operation of `multiplying together' and the associativity expresses the fact that it is immaterial in which order we multiply three or more elements together, the comultiplication $\Delta$ morally corresponds to the operation of `decomposing', and the coassociativity asserts that the order in which this operation is iterated is irrelevant. To be more concrete, take an element $a\in\cC$, and consider its coproduct, schematically written as
\beq\label{eq:Delta(a)}
\Delta(a) = \sum_i a_i^{(1)}\otimes a_i^{(2)}\,,
\eeq
for some $a_i^{(j)}\in\cC$. By acting with $\Delta$ on $a$, we have decomposed it into (a combination of) two pieces. We can iterate this process, and decompose Eq.~\eqref{eq:Delta(a)} further into (a combination of ) three pieces. At this stage, we have the choice to decompose either $a_i^{(1)}$ or $a_i^{(2)}$,
\beq
\Delta(a_i^{(1)}) = \sum_j a_{i,j}^{(1,1)}\otimes a_{i,j}^{(1,2)} {\rm~~and~~}\Delta(a_i^{(2)}) = \sum_j a_{i,j}^{(2,1)}\otimes a_{i,j}^{(2,2)}
\eeq
In the first case we arrive at
\beq\label{eq:cut1}
(\Delta\otimes\id)\Delta(a) = \sum_{i,j} a_{i,j}^{(1,1)}\otimes a_{i,j}^{(1,2)}\otimes a_i^{(2)}\,,
\eeq
while in the second case we arrive at
\beq\label{eq:cut2}
(\id\otimes\Delta)\Delta(a) = \sum_{i,j} a_i^{(1)}\otimes a_{i,j}^{(2,1)}\otimes a_{i,j}^{(2,2)}\,.
\eeq
While Eqs.~\eqref{eq:cut1} and~\eqref{eq:cut2} are in general different, coassociativity asserts that the two expressions must be equal
\beq
(\Delta\otimes\id)\Delta(a) =\sum_{i,j} a_{i,j}^{(1,1)}\otimes a_{i,j}^{(1,2)}\otimes a_i^{(2)} = \sum_{i,j} a_i^{(1)}\otimes a_{i,j}^{(2,1)}\otimes a_{i,j}^{(2,2)} = (\id\otimes\Delta)\Delta(a)\,.
\eeq
As a consequence, there is an essentially unique way to iterate the coproduct,
\beq
\cH\stackrel{\Delta}{\longrightarrow}\cH\otimes\cH
\stackrel{\Delta\otimes\id}{\longrightarrow}\cH\otimes\cH\otimes\cH\stackrel{\Delta\otimes\id\otimes\id}{\longrightarrow}\ldots
\eeq

\subsection{Bialgebras and Hopf algebras}
We are now only one step away from defining a Hopf algebra. First, a \emph{bialgebra} is an algebra that is at the same time a coalgebra, \emph{i.e.}, a vector space equipped with both a multiplication $\mu$ and a comultiplication $\Delta$\footnote{We will from now always tacitly assume that all (co)multiplications are (co)associative.}. We emphasize that in this setting $\mu$ and $\Delta$ are independent and in general not hermitian conjugate to each other. Furthermore, we require the multiplication and the comultiplication to be compatible with each other, \emph{i.e.}, the coproduct of a product equals the product of the coproducts (in other words, $\Delta$ is an algebra homomorphism),
\beq\label{eq:Delta_homomorph}
\Delta(a\cdot b) = \Delta(a)\cdot \Delta(b)\,,
\eeq
where the multiplication in the right-hand side is taken in each factor of the tensor product separately,
\beq
(a_1\otimes a_2)\cdot(b_1\otimes b_2) \equiv (a_1\cdot b_1)\otimes (a_2\cdot b_2)\,.
\eeq

A Hopf algebra $\cH$ is a bialgebra equipped with an additional structure, the so-called antipode $S:\cH\to \cH$ satisfying the properties
\beq
S(a\cdot b) = S(b)\cdot S(a){\rm~~and~~} \mu(\id\otimes S)\Delta = \mu(S\otimes \id)\Delta = 0\,.
\eeq
As in the rest of this paper we do not make explicit use of the antipode, we do not elaborate on it any further.

Let us conclude this section by introducing some notations that will be useful  in subsequent sections. 
Consider a Hopf algebra $\cH$ with coproduct $\Delta$, and assume that $\cH$ is graded (as will be the case for the multiple polylogarithms),
\beq
\cH = \bigoplus_{n=0}^\infty\cH_n\,.
\eeq
If the coproduct respects the weight, we can decompose the action of the coproduct according to
\beq
\cH_n \stackrel{\Delta}{\longrightarrow} \bigoplus_{p+q=n}\cH_p\otimes\cH_q\,.
\eeq
We can then write the action of $\Delta$ on $\cH_n$ as
\beq
\Delta= \sum_{p+q=n}\Delta_{p,q}\,,
\eeq
where $\Delta_{p,q}$ is the part of the coproduct that takes values in $\cH_p\otimes\cH_q$. In a similar way we define $\Delta_{p,q,\ldots,r}$ as the component of the iterated coproduct that takes values in $\cH_p\otimes\cH_q\otimes \ldots\otimes\cH_r$. Finally, it is sometimes useful to define the \emph{reduced} coproduct $\Delta'$ via
\beq
\Delta(a)= 1\otimes a+a\otimes 1 + \Delta'(a)\,.
\eeq
An element $a\in\cH$ such that $\Delta'(a)=0$ is called a \emph{primitive element} of $\cH$.


\section{The multiple polylogarithm Hopf algebra}
\label{sec:Hopf_polylog}
In this section we apply the algebraic concepts of the previous section to multiple polylogarithms. As a result, we obtain a framework that contains the symbol in a certain limit, but is more general and incorporates, in particular, the $\zeta$ values. 

As a starting point, let us denote by $\cH$ the algebra formed by the multiple polylogarithms equipped with the shuffle product. We already know that $\cH$ is graded by the weight of the polylogarithms. In Ref.~\cite{Goncharov-Galois} Goncharov showed that $\cH$ can be equipped with a coproduct which turns it into a Hopf algebra. The coproduct on multiple polylogarithms is given by~\cite{Goncharov-Galois}
\beq\bsp\label{eq:coproduct}
\Delta&(I(a_0;a_1,\ldots,a_n;a_{n+1})) \\
&= \sum_{0=i_1<i_2<\ldots<i_{k}<i_{k+1}=n} I(a_0;a_{i_1},\ldots,a_{i_k};a_{n+1})\otimes\Bigg[\prod_{p=0}^kI(a_{i_p};a_{i_p+1},\ldots,a_{i_{p+1}-1};a_{i_{p+1}})\Bigg]\,.
\esp\eeq
The fact that Eq.~\eqref{eq:coproduct} defines a genuine coproduct, \emph{i.e.}, that $\Delta$ is coassociative, Eq.~\eqref{eq:coassociativity}, and an algebra homomorphism, Eq.~\eqref{eq:Delta_homomorph}, is a non-trivial statement. In addition, Eq.~\eqref{eq:coproduct} preserves the weight, \emph{i.e.}, the sum of the weights in each term is equal to $n$. We stress that Eq.~\eqref{eq:coproduct} is strictly speaking only valid when all the $a_i$'s are generic, \emph{i.e.}, non zero and mutually different. 
The definition of the coproduct in the non-generic case involves several technical steps that do not add anything new to the discussion in the main text of the paper, and we refer to Appendix~\ref{app:coproduct} or to Refs.~\cite{Goncharov-Galois,Goncharov_Mixed_Tate} for the definition of the coproduct in the non-generic case. Let us quote here only the explicit formulas for the coproducts for the ordinary logarithm and the classical polylogarithm,
\beq\bsp\label{eq:coproduct_classical}
\Delta(\ln z) &\,= 1\otimes \ln z + \ln z\otimes 1\,,\\
\Delta(\li_n(z)) &\, = 1\otimes \li_n(z) + \li_n(z)\otimes 1 + \sum_{k=1}^{n-1}\li_{n-k}(z)\otimes {\ln^kz\over k!}\,.
\esp\eeq
Eq.~\eqref{eq:coproduct_classical} is enough to compute the coproduct of any expression made out of ordinary logarithms and classical polylogarithms only. Indeed, we can use Eq.~\eqref{eq:Delta_homomorph} to obtain for example,
\beq\bsp
\Delta(\ln x\ln y) &\,= \Delta(\ln x)\,\Delta(\ln y) = [1\otimes \ln x+\ln x\otimes 1]\,[1\otimes \ln y+\ln y\otimes 1]\\
&\, = 1\otimes(\ln x\,\ln y) + \ln x\otimes \ln y + \ln y\otimes \ln x + (\ln x\,\ln y)\otimes 1\,.
\esp\eeq
Furthermore, it is easy to prove the following result,
 \beq
 \Delta(\ln^nz) = \sum_{k=0}^n\binom{n}{k}\,\ln^kz\otimes \ln^{n-k}z\,.
 \eeq

The coproduct can be used to simplify expressions involving polylogarithms in the same way as the symbol. Indeed, suppose that we have two expression $F_w$ and $G_w$ of weight $w$ that are equal (modulo functional equations). Then it is clear that also their coproducts must be equal,
\beq
\Delta(F_w) = \Delta(G_w)\,,
\eeq
 and also
\beq\label{eq:reduced_coproduct}
\Delta'(F_w) = \Delta'(G_w)\,.
\eeq 
It is important to note that Eq.~\eqref{eq:reduced_coproduct} only involves polylogarithms of weight $w'<w$. As a consequence, it is enough to know the functional equations of lower weight in order to check the equality. These functional equations of lower weight might themselves still be complicated or unknown, so we have apparently not gained anything. In such a scenario we can iterate the procedure by applying the coproduct again to one of the factors in the tensor product, and the coassociativity of the coproduct ensures that this iteration is unique. In this way we obtain a whole tower of expressions, which at each stage involve only transcendental functions of lower weight,
\beq
F_w = G_w \to \Delta(F_w) = \Delta(G_w) \to (\id\otimes\Delta)\Delta(F_w) = (\id\otimes\Delta)\Delta(G_w)
\to \ldots
\eeq
As an example, in the case of a function of weight four, we obtain the following identities,
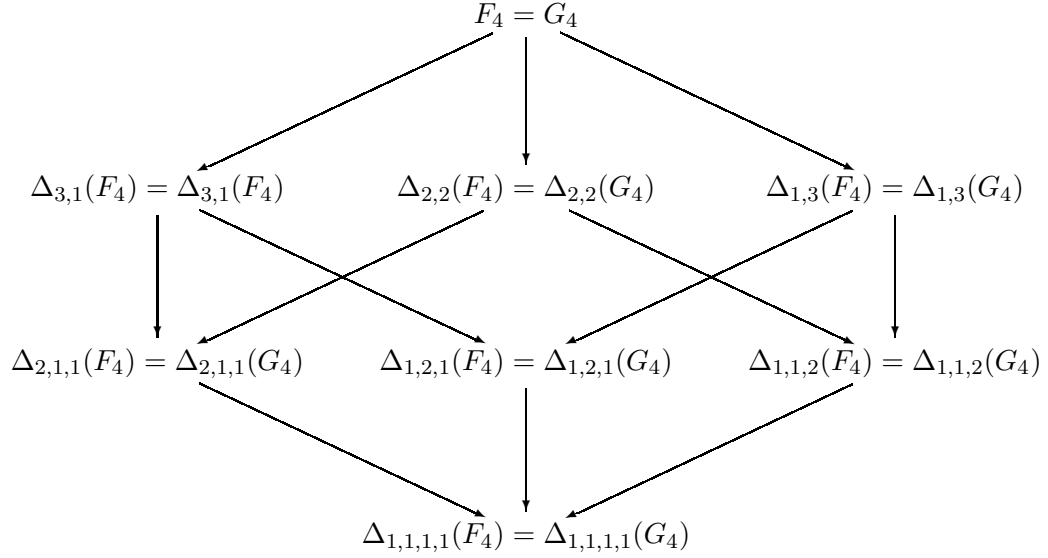
\begin{center}
\beq\nonumber
\begin{diagram}
&&F_4=G_4 && \\
&\ldTo&\dTo&\rdTo&\\
\Delta_{3,1}(F_4) = \Delta_{3,1}(F_4) & \qquad& \Delta_{2,2}(F_4) = \Delta_{2,2}(G_4) &\qquad& \Delta_{1,3}(F_4) = \Delta_{1,3}(G_4)\\
\dTo&\ldTo\rdTo&&\rdTo\ldTo&\dTo\\
\Delta_{2,1,1}(F_4) = \Delta_{2,1,1}(G_4) & \qquad& \Delta_{1,2,1}(F_4) = \Delta_{1,2,1}(G_4) &\qquad& \Delta_{1,1,2}(F_4) = \Delta_{1,1,2}(G_4)\\
&\rdTo&\dTo&\ldTo&\\
&&\Delta_{1,1,1,1}(F_4) = \Delta_{1,1,1,1}(G_4)
\end{diagram}
\eeq
\end{center}
In the extreme case where we go down to $\Delta_{1,\ldots,1}$, we have decomposed a weight $w$ polylogarithm into a tensor of rank $w$ made out of polylogarithms of weight one, \emph{i.e.}, ordinary logarithms, for which all the functional equations are known. It can be shown that in this way we obtain precisely the symbol of the function (up to some technical detail that will be discussed later). In other words, the symbol is nothing but the maximal iteration of the coproduct.
Besides providing a precise definition of the symbol, this approach also shows why the symbol alone is insufficient to determine the function completely. Indeed, requiring two expressions to have the same symbol is equivalent to require that they give the same result when acted upon with $\Delta_{1,\ldots,1}$. While this approach has the obvious advantage that it reduces the problem to the sole application of functional equations for ordinary logarithms, it does in general not imply the equality of the other components of the coproduct. The information on the terms that are missed by the symbol  is nevertheless contained in these other components (at least to some extent). 
To see how this works, and to see how the $\zeta$ values that were missed by the symbol arise in the other components, we first need to overcome some technical obstacles that we will discuss now.

\subsection{The multiple $\zeta$ value Hopf algebra}
In the previous section we argued that the coproduct provides a more general `calculus' that contains the symbol in some limit and that cures most of the unwanted features of the symbol.
However, a consistent `extended symbol calculus' should be compatible with specializations of the arguments.

Multiple $\zeta$ values are, by definition, the values of the multiple polylogarithms (in the series representation) with all arguments equal to unity. It is easy to check that they form in fact a sub-Hopf algebra $\cZ$ of the multiple polylogarithm Hopf algebra $\cH$. From Eq.~\eqref{eq:coproduct_classical} one immediately sees that the coproduct of $\zeta$ values of depth one is given by,
\beq\label{eq:zeta_primitive}
\Delta(\zeta_n) = \Delta(\li_{n}(1)) = 1\otimes\zeta_n + \zeta_n\otimes 1\,,
\eeq
\emph{i.e.}, $\zeta$ value of depth one are primitive elements in $\cZ$, and thus in $\cH$.

At this point we have to face a subtle problem for the even $\zeta$ values. We know that the even $\zeta$ values are not independent, but they are all proportional to powers of $\zeta_{2}$, \emph{e.g.},
\beq
\zeta_4 = {1\over 15}\zeta_2^2\,.
\eeq
Thus,
\beq\bsp
\Delta(\zeta_4) &\,= {1\over 15}\Delta(\zeta_2)^2 = {1\over 15}[1\otimes \zeta_2+\zeta_2\otimes 1]^2={1\over 15}[ 1\otimes \zeta_2^2 + \zeta_2^2\otimes 1 + 2\zeta_2\otimes \zeta_2]\,,
\esp\eeq
and so there is a contradiction with Eq.~\eqref{eq:zeta_primitive}, unless `$\zeta_2 = 0$', \emph{i.e.}, unless we work modulo $\zeta_2$, 
\beq
\Delta(\zeta_2) = 0\,.
\eeq
As a consequence, we lose all information on the terms proportional to $\pi^2$ in the coproduct. 
Hence, if this was the case we would not have gained anything over the naive symbol approach.

In Ref.~\cite{Brown:2011ik} Brown argues that instead of defining the coproduct of $\zeta_2$ to be zero, it is consistent to define
\beq\label{Brown_comodule}
\Delta(\zeta_2) = \zeta_2\otimes 1\,,
\eeq
and more generally
\beq
\Delta(\zeta_{2n}) = \zeta_{2n}\otimes 1\,.
\eeq
This definition obviously solves the problem we had before, because
\beq\bsp
\Delta(\zeta_4) &\,= {1\over 15}\Delta(\zeta_2)^2 = {1\over 15}[\zeta_2\otimes 1]^2={1\over 15}\,\zeta_2^2\otimes 1 = \zeta_4\otimes 1\,.
\esp\eeq

Even though Eq.~\eqref{Brown_comodule} was introduced in Ref.~\cite{Brown:2011ik} in the context of multiple $\zeta$ values, we argue that it equally well holds in more general situations. Moreover, we conjecture that Eq.~\eqref{Brown_comodule} can be extended to
\beq\label{my_conjecture}
\Delta(\pi) = \pi\otimes 1\,.
\eeq
This definition is obviously consistent with Eq.~\eqref{Brown_comodule}. In addition, it allows to extend the coproduct to include the $i\pi$ terms in a consistent way. A word of caution is however in order: due to the monodromy of the logarithm, we should define $2i\,\pi\otimes x = 0$, $\forall x$, and thus also $4\,\pi^2\otimes x= 0$. In practice though, we observed that we never have to worry about the monodromy of the logarithm in physical applications. Indeed, in a physical computation the Riemann sheets of the logarithms are fixed, \emph{e.g.}, by assigning a small imaginary part to $x$, such that
\beq
\ln(x+i\delta\varepsilon) = \ln|x| +\delta\,i\pi\,\theta(-x)\,.
\eeq

The definition~\eqref{my_conjecture} changes the coproduct from being a map
\beq
\Delta:\cH \to \cH \otimes \cH\,,
\eeq
to a map
\beq
\Delta:\cH \to \cH \otimes \cH^\pi\,,
\eeq
where $\cH^\pi$ is the quotient of $\cH$ by the (two-sided) ideal\footnote{We recall that an ideal $\cI$ in a commutative algebra $\cA$ is an additive subgroup such that $\forall a\in \cA$ and $\forall r\in\cI$, we have $a\cdot r\in \cI$.} generated by $\pi$.
The iterated coproduct then takes the form
\beq
\cH \stackrel{\Delta}{\longrightarrow} \cH\otimes\cH^\pi \stackrel{\Delta\otimes\id}{\longrightarrow}
\cH\otimes\cH^\pi\otimes\cH^\pi \stackrel{\Delta\otimes\id\otimes\id}{\longrightarrow}\ldots\,.
\eeq
Loosely speaking, this means that we drop all powers of $\pi$ in all factors of the coproduct, except for the first one.

We are now able to state our main conjecture. We conjecture that using the coproduct together with the definition~\eqref{my_conjecture} we obtain an extension of the symbol calculus that takes into account the $\zeta$ values as well. More precisely, if we have a function $F_w$ of weight $w$ and if we can find a (simpler) function $G_w$ such that
\beq
\Delta'(F_w) = \Delta'(G_w)\,,
\eeq
then
\beq\label{eq:ansatz}
F_w = G_w + \sum_ic_i\, P_{w,i}\,,
\eeq
where the sum runs over all primitive elements of weight $w$ of $\cH$ for some (rational) coefficients $c_i$. We see from Eq.~\eqref{eq:ansatz} that the reduced coproduct does still not entirely fix the function. A similar observation was made in Ref.~\cite{Brown:2011ik} in the case of multiple $\zeta$ values. In practice, the primitive elements turn out to be constants of a given weight, \emph{e.g.},
\begin{enumerate}
\item powers of $\pi$,
\item $\zeta$ values of depth one, $\zeta_n$,
\item Clausen values at the roots of unity, 
\beq
\textrm{Cl}_n\left({k\pi\over N}\right) = \frak{R}_n\,\li_n\left(e^{ik\pi/N}\right)\,,
\eeq
where $\frak{R}_n$ denotes the real part for $n$ even and the imaginary part for $n$ odd.
\end{enumerate}
Even though the function is not entirely fixed, 
we believe that this approach constitutes an important improvement over the pure symbol approach. Indeed, while a pure symbol approach misses for example all functions multiplied by multiple $\zeta$ values, only very few constants are left undetermined by the coproduct. The undetermined constants can easily be fixed by, \emph{e.g.}, comparing to numerical values at a few points, requiring the function to have the right limits, \emph{etc.} While we fall short of a full proof of our conjecture, we have checked the consistency of our `extended symbol calculus' by applying it to hundreds of functional equations among multiple polylogarithms. A selection of results will be shown as an illustration in Sections~\ref{sec:examples} and \ref{sec:Hggg}.

Let us conclude this section by discussing how differential and monodromy operators act on the coproduct, \emph{i.e.}, how to generalize the relations~\eqref{eq:symbol_derivative} and~\eqref{eq:symbol_monodromy} to our framework.
We conjecture that
\beq\bsp\label{eq:diff_mono}
\Delta\left({\partial\over\partial x_k}F_w\right) &\,= \left(\id\otimes{\partial\over\partial x_k}\right)\,\Delta(F_w)\,,\\
\Delta\left(\cM_{x_k=a}F_w\right) &\,= \left(\cM_{x_k=a}\otimes\id\right)\,\Delta(F_w)\,,
\esp\eeq
\emph{i.e.}, differential operators only act in the last component of the coproduct, while monodromy operators only act in the first component. Note that the same statement is true for the iterated coproduct. While we fall short of a full proof of Eq.~\eqref{eq:diff_mono}, we were able to check our claim in the special case where $F_w$ is a multiple polylogarithm with generic arguments. The proofs of these special cases are presented in Appendix~\ref{app:proofs}.

\subsection{Relationship between the coproduct and the symbol}
In this section we briefly discuss the relationship between the coproduct and the symbol. While it is possible to proof in general that the combinatorics of the maximal iteration of the coproduct on multiple polylogarithms matches precisely the combinatorics of the maximal dissections of the rooted and decorated polygon associated to a polylogarithm~\cite{GanglRhodes}, we do not present a firm proof in this paper, but merely state the observation that this correspondence holds in the all cases we have considered. We only motivate the relationship by analyzing how the coproduct behaves under differentiation. If $F_w$ is a transcendental function of weight $w$, then without loss of generality we can write its iterated coproduct in the form
\beq
\Delta_{1,\ldots,1}(F_w) = \sum_i\Delta_{1,\ldots,1}(F_{i,w-1})\otimes\ln R_i\,.
\eeq
If we now act with $(\id\otimes d)$ on this expression, we obtain, using Eq.~\eqref{eq:diff_mono},
\beq\label{eq:eq}
\Delta_{1,\ldots,1}(dF_w) = \sum_i\Delta_{1,\ldots,1}(F_{i,w-1})\otimes d\ln R_i\,,
\eeq
\emph{i.e.}, we obtain an expression which is dual to the differential equation~\eqref{eq:diff_eq_generic} defining the symbol. We emphasize that we claim in no way that this provides a proof of the fact that the maximal iteration contains the symbol, but we hope that it provides a feeling to the reader why this relationship is true.
Note however that the symbol is not exactly equal to $\Delta_{1,\ldots,1}$. Indeed, the symbol does not contain any information about terms proportional to $i\pi$, whereas these terms are incorporated into $\Delta_{1,\ldots,1}$ through Eq.~\eqref{my_conjecture}. In other words, the correct relationship between the symbol and the maximal iteration of the coproduct reads
\beq\label{eq:symb_delta}
\cS \equiv \Delta_{1,\ldots,1} \textrm{ mod }\pi\,.
\eeq

Even though we did not provide a proof of Eq.~\eqref{eq:symb_delta}, we can show that it is consistent with our general knowledge about symbols:
\begin{enumerate}
\item the fact that each entry in a symbol is additive like a logarithm, Eq.~\eqref{eq:symbol_additivity}, is a consequence of the fact that each factor in the tensor product on the left-hand side of Eq.~\eqref{eq:symb_delta} \emph{is} a logarithm.
\item the derivative identity in Eq.~\eqref{eq:diff_mono} reduces to Eq.~\eqref{eq:symbol_derivative} if we restrict ourselves to the maximal iteration of the coproduct. This is obvious from Eq.~\eqref{eq:eq}.
\item similarly, the monodromy identity in Eq.~\eqref{eq:diff_mono} reduces to the corresponding identity for the symbol, Eq.~\eqref{eq:symbol_monodromy}.
\end{enumerate}


\section{Examples}
\label{sec:examples}
In this section we present some simple examples of how the coproduct can be used to simplify expressions involving multiple polylogarithms. The examples in this section do not provide any new results, but they are simple enough so that all the steps can be carried out by hand. They are therefore rather meant to illustrate how to use the coproduct in practise to perform computations.

\subsection{Inversion relations}
We start by considering inversion relations for classical polylogarithms. Throughout this section we assume that $x$ is a real positive variable to which we assign a small positive imaginary part.

We proceed in a bootstrap and build up the inversion relations by a recursion in the weight. 
For the classical polylogarithm of weight 1, the inversion relation is easy to obtain, 
\beq\label{eq:inversion_Li1}
\li_1\left({1\over x}\right) = -\ln\left(1-{1\over x}\right) = -\ln(1-x) +\ln(-x) = -\ln(1-x) + \ln x - i\pi\,.
\eeq
In order to obtain the inversion relation for weight 2, we act with $\Delta_{1,1}$ on $\li_2(1/x)$ and insert the inversion relation for $\li_1(1/x)$,
\beq\bsp
\Delta_{1,1}\left[\li_{2}\left({1\over x}\right)\right] &\,= -\ln\left(1-{1\over x}\right) \otimes \ln\left({1\over x}\right) \\
&\, = \ln(1-x)\otimes\ln x - \ln x\otimes \ln x +i\pi\otimes \ln x\\
 &\,=\Delta_{1,1}\Big[-\li_{2}(x) -{1\over 2}\ln^2x + i\pi\ln x\Big]\,.
\esp\eeq
Following our conjecture, we conclude that the arguments on the left and right-hand sides are equal modulo primitive elements of weight two. We thus make the ansatz,
\beq
\li_{2}\left({1\over x}\right) = -\li_{2}(x) -{1\over 2}\ln^2x + i\pi\ln x + c\,\pi^2\,,
\eeq
for some rational number $c$. Specializing to $x=1$, we immediately obtain $c=1/3$, which is indeed the correct inversion relation for $\li_2$. We emphasize at this stage the importance of the definition~\eqref{my_conjecture}. 

Moving on to weight 3, we act with $\Delta_{1,1,1}$ on $\li_{3}(1/x)$ and obtain
\beq\bsp\label{eq:temp}
\Delta_{1,1,1}\left[\li_{3}\left({1\over x}\right)\right] &\,= -\ln\left(1-{1\over x}\right) \otimes \ln\left({1\over x}\right)\otimes \ln\left({1\over x}\right) \\
&\,= -\ln(1-x)\otimes\ln x\otimes\ln x + \ln x\otimes\ln x\otimes\ln x - i\pi\otimes\ln x\otimes \ln x\\
&\, = \Delta_{1,1,1}\Big[\li_{3}(x) +{1\over 6}\ln^3x - {i\pi\over 2}\ln^2 x\Big]\,.
\esp\eeq
Eq.~\eqref{eq:temp} is not yet the correct inversion relation for $\li_3$.
After subtracting the terms we have found in Eq.~\eqref{eq:temp}, we look at the image of the difference under $\Delta_{2,1}$ or $\Delta_{1,2}$. As an example, we obtain
\beq\bsp
\Delta&_{1,2}\Bigg[\li_{3}\left({1\over x}\right) - \Big(\li_{3}(x) +{1\over 6}\ln^3x - {i\pi\over 2}\ln^2 x\Big)\Bigg] \\ 
&= -{1\over 2}\ln\left(1-{1\over x}\right)\otimes\ln^2\left({1\over x}\right) + {1\over2}\ln(1-x)\otimes\ln^2 x-{1\over2}\ln x\otimes \ln^2x + {i\pi\over 2}\otimes \ln^2x\\
& = 0\,.
\esp\eeq
We see that acting with $\Delta_{1,2}$ does not provide any new information. This is not surprising, as the missing terms are of the form $\pi^2\ln x$, and $\Delta_{1,2}(\pi^2\ln x) = 0$. Indeed, acting with $\Delta_{2,1}$ and using the inversion relation for $\li_2$, we obtain new non-trivial information,
\beq\bsp
\Delta&_{2,1}\Bigg[\li_{3}\left({1\over x}\right) - \Big(\li_{3}(x) +{1\over 6}\ln^3x - {i\pi\over 2}\ln^2 x\Big)\Bigg] \\ 
&= \li_{2}\left({1\over x}\right)\otimes\ln\left({1\over x}\right) -\li_{2}(x)\otimes\ln x-{1\over2}\ln^2 x\otimes \ln x + {i}\,(\pi\ln x)\otimes \ln x\\
& = -\Big[-\li_{2}(x)-{1\over2}\ln^2x + i\pi\ln x + {\pi^2\over 3}\Big]\otimes \ln x\\
&\,\quad -\li_{2}(x)\otimes\ln x-{1\over2}\ln^2 x\otimes \ln x + i\,(\pi\ln x)\otimes \ln x\\
&= -{1\over 3}\,\pi^2\otimes\ln x\\
& = \Delta_{2,1}\Big(-{\pi^2\over 3}\ln x\Big)\,.
\esp\eeq
Thus,
\beq
\li_{3}\left({1\over x}\right) = \li_{3}(x) +{1\over 6}\ln^3x - {i\pi\over 2}\ln^2 x - {\pi^2\over 3}\ln x + \alpha\zeta_3 + \beta\,i\pi^3\,.
\eeq
Specializing to $x=1$ gives $\alpha = \beta = 0$, which is indeed the correct inversion relation for $\li_3$. 
Proceeding in exactly the same way, we can now derive the inversion relations for all the classical polylogarithms.

\subsection{Special values in $x=1/2$}
As a second example we consider the special values of some harmonic polylogarithms when the argument is equal to 1/2. In many cases these values are expressible through $\zeta$ values, $\ln 2$ and $\li_n\left({1\over 2}\right)$, for $n\ge 4$. It is however impossible to obtain these relations using symbols alone, because
\beq
\cS\left[H\left(a_1,\ldots,a_n;{1\over2}\right)\right] = (-1)^p\,2\otimes\ldots\otimes2 = \cS\left[{(-1)^p\over n!} \ln^n2\right]\,,
\eeq
where $a_i\in\{0,1\}$ and $p$ is equal to the number of $a_i$'s equal to zero. As a consequence, a pure symbol approach only provides trivial and misleading information, because we always obtain a symbol corresponding to powers of $\ln 2$.
In the following we show that using the coproduct approach we can do better and entirely fix the values in $x=1/2$, up to primitive elements of a given weight $n$ (in the present case we only need to consider $\zeta_n$).

We again proceed in a bootstrap and start from weight 2. We obtain
\beq
\Delta_{1,1}\left[\li_{2}\left({1\over2}\right)\right]  =  -\ln\left(1-{1\over 2}\right) \otimes\ln\left({1\over2}\right) = -\ln2\otimes\ln 2 = -{1\over2}\,\Delta_{1,1}(\ln^22)\,.
\eeq
Thus,
\beq
\li_{2}\left({1\over2}\right) = -{1\over 2}\ln^22 + c\pi^2\,,
\eeq
for some rational number $c$ that cannot be fixed from the coproduct. Hence, at this stage we need to resort to numerics,
\beq
\li_{2}\left({1\over2}\right) + {1\over 2}\ln^22 = 0.82246703342411321824\ldots = {\pi^2\over 12}\,.
\eeq
It appears at this stage that we have not gained anything over the pure symbol approach. In fact, in this case the coproduct only becomes more powerful than the pure symbol approach starting from weight three. Acting with $\Delta_{1,1,1}$ on $\li_{3}\left({1\over2}\right)$ yields
\beq
\Delta_{1,1,1}\left[\li_{3}\left({1\over2}\right)\right] = \ln2\otimes\ln2\otimes\ln 2 = \Delta_{1,1,1}\Bigg[{1\over 6}\ln^32\Bigg]\,.
\eeq
Next, we can look at the (1,2) and (2,1) components of the coproduct,
\beq\bsp
\Delta_{1,2}\Bigg[\li_{3}\left({1\over2}\right)-{1\over 6}\ln^32\Bigg] &\,= 0\,,\\
\Delta_{2,1}\Bigg[\li_{3}\left({1\over2}\right)-{1\over 6}\ln^32\Bigg] &\,=-\li_{2}\left({1\over2}\right)\otimes \ln2 - {1\over 2} \ln^22\otimes \ln 2\\
&\,=-{1\over 12}\pi^2\otimes\ln 2 = \Delta_{2,1}\Bigg[-{\pi^2\over 12}\ln 2\Bigg]
\esp\eeq
Thus, we obtain
\beq
\li_{3}\left({1\over2}\right) = {1\over 6}\ln^32-{\pi^2\over 12}\ln 2 + \alpha\zeta_3 + \beta i\pi^3\,.
\eeq
$\beta$ is obviously zero, because $\li_{3}\left({1\over2}\right)$ is real. Furthermore, from numerics we obtain
\beq
\li_{3}\left({1\over2}\right) - \Bigg[{1\over 6}\ln^32-{\pi^2\over 12}\ln 2\Bigg] = 1.0517997902646449997\ldots = {7\over8}\,\zeta_3\,.
\eeq

We could now be tempted to apply the same procedure to $\li_{4}\left({1\over2}\right)$ and express it in the form
\beq\label{eq:li4(1/2)}
\li_{4}\left({1\over2}\right) = c_1\,\ln^42 + c_2\,\pi^2\ln^22 + c_3\,\zeta_3\ln2+c_4\,\pi^4\,.
\eeq
However, no such formula is currently known, and it is commonly believed that, starting from $n=4$, $\li_{n}\left({1\over2}\right)$ defines a genuinely new transcendental number. If our `extended symbol calculus' is consistent, it should lead us to the same conclusion, \emph{i.e.},  that an ansatz of the form~\eqref{eq:li4(1/2)} is excluded. To see why this is indeed the case, we start from Eq.~\eqref{eq:coproduct_classical} and we write
\beq\label{eq:lin(1/2)}
\Delta'\Bigg[\li_{n}\left({1\over2}\right)\Bigg] = \sum_{k=1}^{n-1}{(-1)^k\over k!}\,
\li_{n-k}\left({1\over2}\right)\otimes\ln^k2\,.
\eeq
We see that the second factor in the reduced coproduct only involves powers of $\ln 2$. An ansatz made out of combinations of powers of $\ln2$ and $\zeta$ values however inevitably leads to terms in the coproduct that have a $\zeta$ value in the second factor, \emph{e.g.},
\beq
\Delta(\zeta_m\ln^k2) = \Delta(\zeta_m)\,\Delta(\ln^k2) = \zeta_m\otimes\ln^k2+\ln^k2\otimes\zeta_m+\ldots\,.
\eeq
The only way to make the terms having a $\zeta$ value in the second factor vanish is to assume that $m$ is even, because Eq.~\eqref{my_conjecture} implies
\beq
\Delta(\pi^m\ln^k2) = \pi^m\otimes\ln^k2+\ldots\,.
\eeq
Thus, we conclude that any ansatz made out of products of powers of $\ln2$ and $\zeta$ values can only involve even $\zeta$ values. This is indeed what happens for $\li_{3}\left({1\over2}\right)$ where the only possibility for a product of the form $\zeta_m\ln^k2$ is $\pi^2\ln2$. Starting from weight four this is excluded, because Eq.~\eqref{eq:lin(1/2)} involves a term
\beq
\li_{3}\left({1\over2}\right)\otimes\ln^{n-3}2 = {7\over8}\,\zeta_3\otimes\ln^{n-3}2+\ldots\,.
\eeq
We therefore arrive at the conclusion that starting from weight four, $\li_{n}\left({1\over2}\right)$ can no longer be expressed through $\zeta$ values and powers of $\ln2$ alone, in agreement with the common belief. We stress the role played in this argument by the special treatment of $\zeta_2$, Eq.~\eqref{Brown_comodule}.

So far we have only considered examples of classical polylogarithms. Let us therefore conclude this section by discussing a less trivial example of weight five, where the full superiority of the coproduct approach over the pure symbol approach is revealed. Consider to this effect the harmonic polylogarithm $H\left(0,1,0,0,1;{1\over2}\right)$. We can make an ansatz for this number in the form
\beq\bsp
T &\,= c_1\,\li_5\left({1\over2}\right) + c_2\,\li_4\left({1\over2}\right)\,\ln2 
+c_3\,\ln^52 \\
&\,+ c_4\,\pi^2\,\ln^32 + c_5\zeta_3\,\ln^22+c_6\,\pi^4\ln 2 + c_7\,\pi^2\,\zeta_3 + c_8\,\zeta_5\,.
\esp\eeq
Our goal is to find rational numbers $c_i$ such that $H\left(0,1,0,0,1;{1\over2}\right) = T$. A pure symbol approach is obviously totally inadequate for this: not only would it be unable to constrain the coefficients multiplying the $\zeta$ values, but it could also not distinguish between the first three terms in the ansatz, thus only providing a single relation between the coefficients $c_1$, $c_2$ and $c_3$.
As we will see in the following, the coproduct approach allows us to fix all the coefficients, except for the coefficient of $\zeta_5$ (which is a primitive element).

We start by computing the maximal iteration of the coproduct,
\beq\bsp
\Delta_{1,1,1,1,1}\left(\overline{T}\right)
&\, = 
\left(-c_1+5 c_2-120 c_3-1\right) \ln2\otimes \ln2\otimes \ln2\otimes \ln2\otimes \ln2\,,
\esp\eeq
where we introduced the shorthand
\beq
\overline{T} = H\left(0,1,0,0,1;{1\over2}\right) - T\,.
\eeq
Equating the right-hand side to zero, we obtain a relation between $c_1$, $c_2$ and $c_3$,
\beq\label{eq:sol}
c_1 = 5 c_2-120 c_3-1\,.
\eeq
We note that this relation is precisely the information we could extract from a pure symbol approach. We can however now go on and compute the components of the coproduct that involve precisely one factor of weight $w>1$. Inserting the solution~\eqref{eq:sol} into the expression for $T$, we obtain
\beq\bsp
\Delta_{1,1,1,2} \left(\overline{T}\right) 
&\, =
 \Delta_{1,1,2,1} \left(\overline{T}\right)
 =
 \Delta_{1,2,1,1} \left(\overline{T}\right)
 =0\,, \\ 
\Delta_{2,1,1,1} \left(\overline{T}\right) 
&\, = \left(\frac{c_2}{6}-10 c_3-6 c_4\right) \pi ^2\otimes \log 2\otimes \log 2\otimes \log 2\,,
\esp\eeq
and so
\beq
c_4 = \frac{c_2}{36}-{5\over 3} c_3\,.
\eeq
Next we investigate the components of the (iterated) coproduct in $\cH\otimes\cH\otimes\cH$,
\beq\bsp
\Delta_{1,2,2} \left(\overline{T}\right)
&\, =
 \Delta_{2,1,2} \left(\overline{T}\right)
 =
 \Delta_{2,2,1} \left(\overline{T}\right) =0\,,
 \esp\eeq
and
\beq\bsp
\Delta_{1,1,3} \left(\overline{T}\right)
&\, = \left(\frac{7}{4}-2 c_5\right) \log 2\otimes \log 2\otimes \zeta_3\,,\\
\Delta_{1,3,1} \left(\overline{T}\right)  
&\, = \left(\frac{7 c_2}{8}-2 c_5-\frac{7}{8}\right) \log 2\otimes \zeta_3\otimes \log2\,,\\
\Delta_{3,1,1} \left(\overline{T}\right) 
&\, = \left(-\frac{21 c_2}{8}+105 c_3-2 c_5\right) \zeta_3\otimes \log 2\otimes \log 2\,,
 \esp\eeq
 and so we can solve for $c_2$, $c_3$ and $c_5$,
 \beq
c_2 = 3 {\rm~~and~~} c_3 = {11\over 120} {\rm~~and~~}c_5 = {7\over 8}\,.
\eeq
For the (3,2) and (2,3) components of the coproduct we obtain
\beq\bsp
\Delta_{3,2} \left(\overline{T}\right) 
&\, = 0 {\rm~~and~~}
\Delta_{2,3} \left(\overline{T}\right)
 = -\left(c_7+\frac{7}{48}\right) \pi ^2\otimes \zeta_3\,,
\esp\eeq
and so $c_7 = -{7\over 48}$.
Similarly the (4,1) and (1,4) components yield
\beq\bsp
\Delta_{1,4} \left(\overline{T}\right) 
&\, = 0 {\rm~~and~~}
\Delta_{4,1}\left(\overline{T}\right)
= \left(\frac{1}{288}-c_6\right) \pi ^4\otimes \log2\,,
\esp\eeq
and so $c_6 = {1\over 288}$.
Finally we arrive at 
\beq\bsp
H\left(0,1,0,0,1;{1\over2}\right) &\,= 3\, \text{Li}_5\left(\frac{1}{2}\right)
+3\, \log 2\, \text{Li}_4\left(\frac{1}{2}\right) +\frac{11}{120}\,\log ^52-\frac{5}{72}\, \pi ^2\, \log ^32\\
&\,+\frac{7}{8}\, \zeta_3\, \log ^22+\frac{1}{288}\, \pi ^4\, \log 2
-\frac{7 }{48}\,\pi ^2\, \zeta_3+ c_8\, \zeta_5\,.
\esp\eeq
As expected, the coproduct allowed us to fix all the coefficients except for $c_8$. Using numerics, we arrive at
\beq
H\left(0,1,0,0,1;{1\over2}\right) - T = -c_8\, \zeta_5-1.3123616901033275\ldots = -c_8\, \zeta_5 -{81\over64}\,\zeta_5\,,
\eeq
and thus $c_8 = -{81\over 64}$.

\section{Amplitudes for H + 3 gluons}
\label{sec:Hggg}
In this section we apply the coproduct to a physical problem, namely the two-loop helicity amplitudes for a Higgs boson plus three gluons in the large top mass limit. In this limit the coupling of a Higgs boson to gluons is described by an effective operator of dimension five,
\beq
{\begin{cal}L\end{cal}}_{eff} = -{\lambda\over4}\,H\,G^a_{\mu\nu}\,G_a^{\mu\nu}\,.
\eeq
The two-loop corrections to the helicity amplitudes for a Higgs boson plus three gluons were computed in Refs.~\cite{Koukoutsakis,Gehrmann:2011aa}, where it was expressed as a complicated combination of two-dimensional harmonic polylogarithms. 
In Ref.~\cite{Brandhuber:2012vm} it was shown that, after subtracting the square of the one-loop amplitude, the symbol of the leading color maximally transcendental part of the two-loop helicity amplitudes is equal to the symbol of the two-loop form factor of three gluons in planar $\cN=4$ Super Yang-Mills. The latter can be expressed in a very compact form involving only classical polylogarithms up to weight four~\cite{Brandhuber:2012vm}. This suggests that the two-loop corrections to the helicity amplitudes for a Higgs boson plus three gluons can be written in a much simpler form without any multiple polylogarithms. However, as the symbol does not fix terms proportional to $\zeta$ values, the symbol alone is insufficient to determine such a simplified form in an easy way. In the following we apply our coproduct approach to rewrite the results of Refs.~\cite{Koukoutsakis,Gehrmann:2011aa} in a compact form, obtaining in this way compact analytical expressions for all helicity amplitudes for a Higgs boson plus three gluons, for both the decay ($H\to ggg$) and the scattering ($gg\to Hg$) regions.

\subsection{The decay region}
We start by investigating the decay region, \emph{i.e.}, the two-loop corrections to the helicity amplitudes for $H\to ggg$. The kinematics is described by three dimensionless ratios,
\beq
x_1 = {s_{12}\over m_H^2}\,,\qquad x_2 = {s_{23}\over m_H^2}\,,\qquad x_3 = {s_{31}\over m_H^2}\,,
\eeq
where $m_H$ denotes the mass of the Higgs boson and $s_{ij} = 2p_ip_j$, with $p_i$ the momenta of the external gluons. These kinematic variables are not independent, but they are constraint by
\beq\label{eq:+++_constraint}
0<x_i<1 {\rm~~and~~} x_1+x_2+x_3=1\,.
\eeq
As a consequence, the amplitude is effectively a function of only two of the three dimensionless ratios. Correspondingly, the result of Ref.~\cite{Gehrmann:2011aa} is expressed in terms of two-dimensional harmonic polylogarithms in $x_2$ and $x_3$. There are two independent helicity configurations for the decay,
\beq
H\to g^+g^+g^+ {\rm~~and~~} H\to g^+g^+g^-\,.
\eeq
In the following we will analyze each configuration separately.

Let us start by analyzing the helicity amplitude where all the final state gluons have a positive helicity. Bose symmetry then implies that the amplitude must be symmetric under a permutation of the external gluons, or, equivalently, it must be totally symmetric in the kinematic variables $x_i$, $i=1,2,3$. The (finite part of the) one-loop correction to the decay can be written as
\beq\label{eq:alpha_1_+++}
\alpha^{(1)} = N_c A^{(1)}_\alpha + \left({N_c\over6} - {11\over12}\,N_f\right)\,B^{(1)}_\alpha+{1\over 6}(N_c-N_f) (x_1\,x_2+x_2\,x_3+x_3\,x_1)\,,
\eeq
with
\beq\bsp
A^{(1)}_\alpha &\,= \frac{\pi ^2}{4}-\frac{1}{2} (\ln x_1\, \ln x_2+\ln x_2\, \ln x_3+\ln x_3\, \ln x_1) - \sum_{i=1}^3\li_2(1-x_i)\,,\\
B^{(1)}_\alpha &\,= \ln(x_1\,x_2\,x_3) - 3\pi i\,.
\esp\eeq
Following Ref.~\cite{Gehrmann:2011aa}, we decompose the (finite part of the) two-loop correction into contributions with different color structures, and we furthermore subtract the square of the finite part of the one-loop amplitude, Eq.~\eqref{eq:alpha_1_+++},
\beq\label{eq:alpha_2_+++}
\alpha^{(2)} = {1\over 2}\Big[\alpha^{(1)}\Big]^2  + N_c^2\,\overline{A}_\alpha^{(2)} + {N_f\over N_c}\,\overline{D}_{\alpha}^{(2)}+N_c\,N_f\,\overline{E}_{\alpha}^{(2)} + N_f^2\,\overline{F}_{\alpha}^{(2)}\,.
\eeq
The coefficients of the different color structures were computed in Ref.~\cite{Gehrmann:2011aa} where they were expressed as a combination of two-dimensional harmonic polylogarithms in $x_2$ and $x_3$. In order to simplify these expressions, we start by computing the symbol of Eq.~\eqref{eq:alpha_2_+++}. It turns out that all the entries in the symbol are drawn from the set
\beq\label{eq:entries}
\{x_1,x_2,x_3,1-x_1,1-x_2,1-x_3\}\,.
\eeq
The weight four part of the symbol satisfies 
\beq
\delta\left[\cS\left(\alpha^{(2)}_{|\textrm{weight 4}}\right)\right] = 0\,,
\eeq
where $\delta(a\otimes b\otimes c\otimes d) = (a\wedge b)\wedge(c\wedge d)$, and the wedge denotes the antisymmetric tensor product, $a\wedge b=a\otimes b- b\otimes a$. It follows then from a conjecture in Ref.~\cite{Goncharov_delta} that $\alpha^{(2)}$ can be expressed in terms of classical polylogarithms only. Similar conclusions were already drawn in Ref.~\cite{Brandhuber:2012vm}.

\begin{table}[!t]
\begin{center}
\begin{tabular}{c|c|c}
\hline\hline
$x_1$ & $1-x_1$ & $1-1/x_1$  \\
$x_2$ & $1-x_2$ & $1-1/x_2$  \\
$x_3$ & $1-x_3$ & $1-1/x_3$ \\
$-{x_1}/{x_2}$ & $x_2/({1-x_3})$ & $x_1/({1-x_3})$\\
$-{x_2}/{x_3}$ & $x_3/({1-x_1})$ & $x_2/({1-x_1})$\\
$-{x_3}/{x_1}$ & $x_1/({1-x_2})$ & $x_3/({1-x_2})$\\
$-{x_1 x_2}/{x_3}$ & $x_3/[{(1-x_1)(1-x_2)}]$ & $x_1x_2/[{(1-x_1)(1-x_2)}]$ \\
$-{x_2 x_3}/{x_1}$ & $x_1/[{(1-x_2)(1-x_3)}]$ & $x_2x_3/[{(1-x_2)(1-x_3)}]$ \\
$-{x_3 x_1}/{x_2}$ & $x_2/[{(1-x_3)(1-x_1)}]$ & $x_3x_1/[{(1-x_3)(1-x_1)}]$ \\
\hline\hline
\end{tabular}
\caption{\label{tab:args}Arguments of classical polylogarithms that can give rise to a symbol with entries drawn from the set in Eq.~\eqref{eq:entries} under the constraint~\eqref{eq:+++_constraint}. Each line shows half an orbit of the $S_3$ action, the second half being obtained by inversion. All these functions are less than unity in the region defined by Eq.~\eqref{eq:+++_constraint}.}
\end{center}
\end{table}

Next, we have to determine the arguments of the polylogarithms that can lead to a symbol with entries drawn from the set~\eqref{eq:entries} under the constraint~\eqref{eq:+++_constraint}. Using the prescription given in Ref.~\cite{Duhr:2011zq}, we find 54 rational functions grouping into 9 orbits of the symmetric group $S_3$ whose action on rational functions $f(x_1,x_2,x_3)$ is generated by\footnote{We stress that this $S_3$ symmetry is \emph{not} identical to the $S_3$ describing the Bose symmetry.}
\beq
f \to 1-f {\rm~~and~~} f\to 1/f\,.
\eeq
The rational functions are summarized in Table~\ref{tab:args}. It is important to note that not all 54 solutions are independent, and in particular we can express half of them in terms of the others by using the inversion relation for the classical polylogarithms,
\beq
\li_n\left({1\over f}\right) = (-1)^{n+1}\,\li_n(f) + \ldots\,.
\eeq
It is then easy to see that it is always possible to choose 27 solutions such that all polylogarithms are real in the region defined by Eq.~\eqref{eq:+++_constraint}.

Next step we write down a combination of (classical) polylogarithms in the arguments shown in Table~\ref{tab:args}. Equating the symbol of $\alpha^{(2)}$ and our ansatz provides a linear system for the coefficients. In the following we only discuss the weight four part of $\overline{A}^{(2)}_\alpha$. All other contributions are similar. In agreement with Ref.~\cite{Brandhuber:2012vm}, we find
\beq\label{eq:symb_+++}
\cS\left(\overline{A}^{(2)}_{\alpha, \textrm{ weight 4}}\right) = \cS\left(\cR^{(2)}_3\right)\,,
\eeq
where $\cR^{(2)}_3$ is the $\cN=4$ form factor remainder function of Ref.~\cite{Brandhuber:2012vm},
\beq\bsp
\cR^{(2)}_3&\,= -\frac{1}{12} \left[\Lambda _4\left(-\frac{x_1 x_2}{x_3}\right)+\Lambda _4\left(-\frac{x_1 x_3}{x_2}\right)+\Lambda_4\left(-\frac{x_2 x_3}{x_1}\right)\right]-2\sum_{i=1}^3\li_4\left(1-{1\over x_i}\right)\\
&\,-{1\over2}\left[\sum_{i=1}^3\li_2\left(1-{1\over x_i}\right)\right]^2-\frac{2}{3} \ln x_1\, \ln x_2\, \ln x_3\, \ln \left(x_1 x_2 x_3\right) +{1\over 16}\ln^4(x_1x_2x_3)\\
&\,+ \frac{1}{3} (\ln x_1\, \ln x_2+\ln x_1\, \ln x_3+\ln x_2\, \ln x_3) \left(\ln^2 x_1+\ln^2 x_2+\ln^2 x_3\right)-\frac{23 \pi ^4}{720}\,,
\esp\eeq
where $\Lambda_n(z)$ denotes Kummer's function,
\beq
\Lambda_n(z) = \int_0^z\rd t\,{\ln^{n-1}|t|\over 1+t} = (n-1)!\sum_{k=0}^{n-1}{(-1)^{n-k}\over k!}\,\ln^k|z|\,\li_{n-k}(z)\,.
\eeq
This result was already obtained in Ref.~\cite{Brandhuber:2012vm}. However, Eq.~\eqref{eq:symb_+++} only holds at the level of the symbol, and it would thus be premature to conclude that the weight four part of $\overline{A}^{(2)}_\alpha$ is equal (at the level of the function) to $\cR^{(2)}_3$. Indeed, acting with $\Delta_{2,1,1}$, we obtain
\beq
\Delta_{2,1,1}\Big[\overline{A}^{(2)}_{\alpha\textrm{, weight 4}} - \cR^{(2)}_3\Big] = - {1\over 6}\, \pi^2\otimes \Delta_{1,1}\Big[A^{(1)}_\alpha \Big]
= \Delta_{2,1,1}\Big[- {\pi^2\over 6} \,  A^{(1)}_\alpha\Big]\,.
\eeq
Continuing this way, we can easily determine the coefficient of $\zeta_3$,
\beq
\Delta_{3,1}\Big[\overline{A}^{(2)}_{\alpha\textrm{, weight 4}} - \cR^{(2)}_3 + {\pi^2\over 6} \,  A^{(1)}_\alpha\Big] =
 -{1\over 4}\,\zeta_3\otimes B^{(1)}_\alpha = \Delta_{3,1}\Big[ -{1\over 4}\,\zeta_3\,B^{(1)}_\alpha\Big]\,.
 \eeq
 Finally, we determine the coefficient of $\pi^4$ by evaluating the function at a single point in phase space,
 \beq
 \overline{A}^{(2)}_{\alpha\textrm{, weight 4}} - \cR^{(2)}_3 + {\pi^2\over 6} \,  A^{(1)}_\alpha + {1\over 4}\,\zeta_3\,B^{(1)}_\alpha
 = -0.03382260105347\ldots=-{\pi^4\over2880}\,.
 \eeq
 Repeating the same steps for all other contributions to Eq.~\eqref{eq:alpha_2_+++}, we arrive at the following expressions for the different color structures contributing to the two-loop amplitude $\alpha^{(2)}$,
\beq\bsp\label{eq:A_alpha_2}
\overline{A}_\alpha^{(2)} &\,= \cR^{(2)}_3 - {\pi^2\over 6} \,  A^{(1)}_\alpha -{1\over 4}\,\zeta_3\,B^{(1)}_\alpha -{\pi^4\over 2880}\\
&\,\frac{11}{6}\Bigg\{\Lambda _3\left(-\frac{x_1 x_3}{x_2}\right)+\Lambda _3\left(-\frac{x_2 x_3}{x_1}\right) + \Lambda _3\left(-\frac{x_1 x_2}{x_3}\right)-\sum_{i=1}^3\li_3\left(1-{1\over x_i}\right)\\
&\, -\Lambda _3\left(-\frac{x_1}{x_2}\right)-\Lambda _3\left(-\frac{x_2}{x_1}\right)-\Lambda _3\left(-\frac{x_1}{x_3}\right)-\Lambda _3\left(-\frac{x_3}{x_1}\right)-\Lambda _3\left(-\frac{x_2}{x_3}\right)-\Lambda _3\left(-\frac{x_3}{x_2}\right)\\
&\,+{1\over 2}\ln(x_1\,x_2\,x_3)\,A^{(1)}_\alpha + {7\over 2}\sum_{i=1}^3\left[\li_2\left(1-{x_i}\right)\,\ln x_i\right]
+\frac{3}{4} \ln x_ 1\, \ln x_ 2\, \ln x_ 3+\frac{1}{6} \ln^3\left(x_1 x_2 x_3\right)\\
&\,-{5\over16}\pi^2\ln(x_1x_2x_3)-{3\over8}\zeta_3+i\pi\,A^{(1)}_\alpha+{i\pi^3\over16}-{1\over 3}\sum_{i=1}^3\ln^3x_i
\Bigg\}\\
&\,+ {1\over 36}\sum_{i=1}^3\Big[{P_1(x_i,x_{i-1},x_{i+1})\over x_{i-1}^2x_{i+1}^2}\,\li_2(1-x_i)+{P_2(x_i,x_{i-1},x_{i+1})\over x_{i}^2}\ln x_{i-1}\ln x_{i+1} +{121\over 4}\ln^2x_i\Bigg]\\
&\,+{P_3(x_1,x_2,x_3)\over 144x_1^2x_2^2x_3^2}\pi^2 - {121\over72}i\pi\ln(x_1x_2x_2) 
+{11\over 36}i\pi\,(x_1x_2 +x_2x_3+x_3x_1) + {185\over 24}i\pi\\
&\,+{1\over72}\sum_{i=1}^3{P_4(x_i,x_{i-1},x_{i+1})\over x_{i-1}x_{i+1}}\ln x_i -\frac{1}{72} \left(x_1 x_2+x_3 x_2+x_1 x_3\right){}^2+\frac{247}{108} \left(x_1 x_2+x_3 x_2+x_1 x_3\right)\\
&\,+\frac{1321}{216}
\,,
\esp\eeq
\beq\bsp
\overline{D}_{\alpha}^{(2)} &\,= -\zeta_3 +{i\pi\over 4}-\frac{1}{6} \left(x_1 x_2+x_3 x_2+x_1 x_3\right)+\frac{67}{48}+{P_5(x_1,x_2,x_3)\over 72x_1^2x_2^2x_3^2}\pi^2 \\
&\,+{1\over 12}
\sum_{i=1}^3\Bigg[{P_6(x_i,x_{i-1},x_{i+1})\over x_{i-1}^2x_{i+1}^2}\,\li_2(1-x_i)+{P_7(x_i,x_{i-1},x_{i+1})\over x_{i}^2}\ln x_{i-1}\ln x_{i+1}\\
&\,+{P_8(x_i,x_{i-1},x_{i+1})\over 2x_{i-1}x_{i+1}}\ln x_{i}\Bigg]
\esp\eeq
\beq\bsp
\overline{E}_{\alpha}^{(2)} &\,= -{i\pi^3\over48}-{i \pi\over 3}\,A^{(1)}_\alpha  -\frac{1}{12} \ln \left(x_1 x_2 x_3\right) (\ln x_1\, \ln x_2+\ln x_1\, \ln x_3+\ln x_2\, \ln x_3) \\
&\,+ {P_{13}(x_1,x_2,x_3)\over432}+\frac{7}{12} \ln x_1\, \ln x_2\, \ln x_3-\frac{5}{48} \pi ^2 \ln \left(x_1 x_2 x_3\right)-\frac{29}{24}\zeta_3\\
&\, + {11\over18}\,i\pi\,\ln(x_1x_2x_3)+{P_{11}(x_1,x_2,x_3)\over288x_1^2x_2^2x_3^2}\,\pi^2+\sum_{i=1}^3\Bigg[\li_3(x_i) - {1\over3} \li_3(1-x_i) \\
&\,+{1\over6} \li_2(1-x_i)\,\ln x_i + {1\over2} \ln(1-x_i)\,\ln^2x_i 
+{1\over6}\ln(x_1x_2x_3)\,\li_2(1-x_i)\\
&\,+{P_9(x_i,x_{i-1},x_{i+1})\over36x_{i-1}^2x_{i+1}^2}\,\li_2(1-x_i)+{P_{10}(x_i,x_{i-1},x_{i+1})\over36x_{i}^2}\,\ln x_{i-1}\,\ln x_{i+1} \\
&\,+ {11\over 36}\ln^2 x_i
+{P_{12}(x_i,x_{i-1},x_{i+1})\over216x_{i-1}x_{i+1}} \,\ln x_i
\Bigg]-\frac{13}{36} i \pi  \left(x_1 x_2+x_3 x_2+x_1 x_3\right)-\frac{71 }{18}i \pi \,,
\esp\eeq
\beq\bsp\label{eq:F_alpha_2}
\overline{F}_{\alpha}^{(2)} &\,= -{i\pi\over18}\,\ln(x_1x_2x_3) - {11\over144}\pi^2 + 
     {1\over36} \sum_{i=1}^3\ln^2 x_i -{5\over54}\, \ln(x_1x_2x_3)+
     {5i\pi\over18}\\
     &\,+{i\pi\over18}\,(x_1x_2+x_2x_3+x_3x_1)+\frac{5}{54} \left(x_1 x_2+x_3 x_2+x_1 x_3\right)\\
     &\,-\frac{1}{72} \left(x_1 x_2+x_3 x_2+x_1 x_3\right)^2-
     {x_1 x_2 x_3\over18} \sum_{i=1}^3{\ln x_i\over x_i}\,,
     \esp\eeq
where $P_i(x,y,z) = P_i(x,z,y)$ are homogeneous polynomials in three variables, 
\beq
\bsp
P_1(x,y,z)&\, = 30 x^2 \left(y^4+z^4\right)-134 y^2 z^2 \left(x^2+y^2+z^2\right)-199 x y^2 z^2 (y+z)\\
&\,+75 x y z \left(x y^2+x z^2+y^3+z^3\right)-268 y^3 z^3\,,\\
P_2(x,y,z)&\, = -134 x^3 (y+z)-67 x^2 \left(x^2+y^2+z^2\right)-65 x^2 y z+75 x y z (y+z)+30 y^2 z^2\,,\\
P_3(x,y,z)&\, =-20 \left(x^4 y^4+x^4 z^4+y^4 z^4\right)-276 x^2 y^2 z^2 (x y+x z+y z)\\
&\,-50 x y z \left(x^3 y^2+x^3 z^2+x^2 y^3+x^2 z^3+y^3 z^2+y^2 z^3\right)\\
&\,-115 x^2 y^2 z^2 \left(x^2+y^2+z^2\right)\,,\\
P_4(x,y,z)&\, =60 x^2 \left(y^2+z^2\right)+9 x^2 y z-78 x y z (y+z)-256 y^2 z^2-117 y z \left(y^2+z^2\right)\,,\\
P_5(x,y,z)&\, =x^4 y^4+x^4 z^4+y^4 z^4-2 x y z \left(x^3 y^2+x^3 z^2+x^2 y^3+x^2 z^3+y^3 z^2+y^2 z^3\right)\,,
\esp\eeq
\beq\bsp
P_6(x,y,z)&\, =-x \left[x \left(y^4+z^4\right)-2 x y z \left(y^2+z^2\right)-2 y z \left(y^3+z^3\right)\right]\,,\\
P_7(x,y,z)&\, = y z (2 x y+2 x z-y z)\,,\\
P_8(x,y,z)&\, = -x \left[2 x \left(y^2+z^2\right)-5 x y z-4 y z (y+z)\right]\,,\\
P_9(x,y,z)&\, = -33 x^2 \left(y^4+z^4\right)+20 y^2 z^2 \left(x^2+y^2+z^2\right)-2 x y^2 z^2 (y+z)\\
&\,-42 x y z \left(x y^2+x z^2+y^3+z^3\right)+40 y^3 z^3\,,\\
P_{10}(x,y,z)&\, =20 x^3 (y+z)+10 x^2 \left(x^2+y^2+z^2\right)-22 x^2 y z-42 x y z (y+z)-33 y^2 z^2\,,\\
P_{11}(x,y,z)&\, = 44 \left(x^4 y^4+x^4 z^4+y^4 z^4\right)+410 x^2 y^2 z^2 (x y+x z+y z)\\
&\,+56 x y z \left(x^3 y^2+x^3 z^2+x^2 y^3+x^2 z^3+y^3 z^2+y^2 z^3\right)\\
&\,+177 x^2 y^2 z^2 \left(x^2+y^2+z^2\right)\,,\\
P_{12}(x,y,z)&\, =-198 x^2 \left(y^2+z^2\right)+35 x^2 y z+142 x y z (y+z)+490 y^2 z^2+206 y z \left(y^2+z^2\right)\,,\\
P_{13}(x,y,z)&\, = -1781 \left(x^4+y^4+z^4\right)-8224 \left(x^3 y+x^3 z+x y^3+x z^3+y^3 z+y z^3\right)\\
&\,-12874 \left(x^2 y^2+x^2 z^2+y^2 z^2\right)-26848 x y z\,.
\esp\eeq
Note that $P_3$, $P_5$, $P_{11}$ and $P_{13}$ are totally symmetric. The expressions for the two-loop corrections to $\alpha^{(2)}$ in Eqs.~(\ref{eq:A_alpha_2} - \ref{eq:F_alpha_2}) only involve classical polylogarithms with a rather simple dependence on the kinematic invariants. In particular, the functional dependence is such that all the polylogarithms are real in the region~\eqref{eq:+++_constraint}. Furthermore, the $S_3$ Bose symmetry of the amplitude if completely manifest in our expressions.

We can of course apply exactly the same procedure to the second helicity configuration, $H\to g^+g^+g^-$. Bose symmetry implies that this amplitude must be symmetric under an exchange of the two positive-helicity gluons, or, equivalently, under an exchange of $x_2$ and $x_3$. 
The one-loop corrections are given by
\beq
\beta^{(1)} = N_c A^{(1)}_\beta + \left({N_c\over6} - {11\over12}\,N_f\right)\,B^{(1)}_\beta+{1\over 6}(N_c-N_f) {x_2x_3\over x_1}\,,
\eeq
with
\beq
A^{(1)}_\beta = A^{(1)}_\alpha {\rm~~and~~} B^{(1)}_\beta = B^{(1)}_\alpha\,.
\eeq
For the two-loop corrections we write
\beq
\beta^{(2)} = {1\over 2}\Big[\beta^{(1)}\Big]^2  + N_c^2\,\overline{A}_\beta^{(2)} + {N_f\over N_c}\,\overline{D}_{\beta}^{(2)}+N_c\,N_f\,\overline{E}_{\beta}^{(2)} + N_f^2\,\overline{F}_{\beta}^{(2)}\,,
\eeq
Repeating the same steps as for the first helicity configuration, we arrive at
\beq\bsp
\overline{A}_\beta^{(2)} &\,= \cR^{(2)}_3 - {\pi^2\over 6} \,  A^{(1)}_\alpha -{1\over 4}\,\zeta_3\,B^{(1)}_\alpha -{\pi^4\over 2880}+{11\over 6}i\pi\,A^{(1)}_\alpha+ {11\over 96}i\pi^3\\
&\,+{11\over 3}\li_{3}(1-x)+{11\over 6}\li_{3}\left(1-{1\over x}\right)-{11\over288}\pi^2\,\ln x_1 -{11\over 12}\ln(x_2x_3)\li_2(1-x_1) \\
&\,-{11\over 36}\ln^3x_1+ {11\over 24}\ln^2x_1\ln(x_2x_3)+ {11\over 24}\ln^2x_2\ln(x_1x_3)+ {11\over 24}\ln^2x_3\ln(x_2x_1)\\
&\,+{11\over 6}\ln x_1\ln x_2\ln x_3 +{Q_7(x_1,x_2,x_3)\over 48x_1^3}\zeta_3
\esp\eeq
\beq\bsp
&\,+
{Q_3(x_1,x_2,x_3)\over 6x_1^3}\li_3(1-x_2) +{Q_3(x_1,x_3,x_2)\over 6x_1^3}\li_3(1-x_3)\\
&\,
+
{Q_4(x_1,x_2,x_3)\over 6x_1^3}\li_3\left(1-{1\over x_2}\right) +{Q_4(x_1,x_3,x_2)\over 6x_1^3}\li_3\left(1-{1\over x_3}\right) \\
&\,
+{1\over 12}\li_2(1-x_2)\Bigg[\left(11-{Q_4(x_1,x_2,x_3)\over x_1^3}\right)\ln x_2 + {Q_5(x_1,x_2,x_3)\over x_1^3}\ln x_3\Bigg] \\
&\,
+{1\over 12}\li_2(1-x_3)\Bigg[\left(11-{Q_4(x_1,x_3,x_2)\over x_1^3}\right)\ln x_3 + {Q_5(x_1,x_3,x_2)\over x_1^3}\ln x_2\Bigg] \\
&\,-{Q_4(x_1,x_2,x_3)\over 36x_1^3}\ln^3 x_2-{Q_4(x_1,x_2,x_3)\over 36x_1^3}\ln^3 x_3\\
&\,+{Q_6(x_1,x_2,x_3)\over 288x_1^3}\pi^2\,\ln x_2+{Q_6(x_1,x_3,x_2)\over 288x_1^3}\pi^2\,\ln x_3\\
&\,+{Q_9(x_1,x_2,x_3)\over 36x_2^2x_3^2}\,\li_2(1-x_1)+{Q_8(x_1,x_2,x_3)\over 36x_1^4x_3^2}\,\li_2(1-x_2)+{Q_8(x_1,x_3,x_2)\over 36x_1^4x_2^2}\,\li_2(1-x_3)\\
&\,-{67\over36}\left(\ln x_1\ln x_2 + \ln x_2\ln x_3 + \ln x_3\ln x_1\right)
+{121\over144}\left(\ln^2x_1+\ln^2x_2+\ln^2x_3\right)\\
&\,+{5\over 12}\ln x_1\left[{x_2(2x_2+5x_3)\over x_3^2}\ln x_2+{x_3(2x_3+5x_2)\over x_2^2}\ln x_3\right] +{Q_{10}(x_1,x_2,x_3)\over144x_1^4x_2^2x_3^2}\pi^2\\
&\,-{x_2x_3\over12x_1^4}[31x_1^2+8x_1(x_2+x_3)-30x_2x_3]\ln x_2\ln x_3 - {121\over72}i\pi\ln(x_1x_2x_3)\\
&\,+{555x_1^2+22x_2x_3\over72x_1^2}i\pi 
+ {Q_2(x_1,x_2,x_3)\over 72x_1^2x_2x_3}\ln x_1\\
&\,
+ {Q_1(x_1,x_2,x_3)\over 24x_1^3x_3(1-x_2)^2}\ln x_2+ {Q_1(x_1,x_3,x_2)\over 24x_1^3x_2(1-x_3)^2}\ln x_3\\
&\,+{745x_1^2 - 198x_1(x_2 + x_3) + 674x_2x_3\over216x_1^2} +
     {4(1+x_1)\over 3(1-x_2)(1-x_3)}- {x_2^2x_3^2\over72x_1^4}\,.
\esp\eeq
\beq\bsp
\overline{D}_\beta^{(2)} &\,= -\zeta_3 -\frac{x_2^4-2 x_3 x_2^3-2 x_3^3 x_2+x_3^4}{12 x_2^2 x_3^2}\text{Li}_2\left(1-x_1\right)+{Q_{25}(x_1,x_2,x_3)\over72x_1^4x_2^2x_3^2}\pi^2\\
&\,
-\frac{\left(1-x_2\right)^2 x_2Q_{24}(x_1,x_2,x_3) }{12 x_1^4 x_3^2}\text{Li}_2\left(1-x_2\right)
-\frac{\left(1-x_3\right)^2 x_3 Q_{24}(x_1,x_3,x_2)}{12 x_1^4 x_2^2}\text{Li}_2\left(1-x_3\right)\\
&\,-\frac{1}{12} \ln x_1\, \left[\frac{x_2 \left(x_2-2 x_3\right)}{x_3^2}\ln x_2+\frac{x_3 \left(x_3-2 x_2\right)}{x_2^2}\ln x_3\right]+{Q_{22}(x_1,x_2,x_3)\over48x_1^2(1-x_2)(1-x_3)}\\
&\,
-\frac{x_2 x_3 \left(6 x_1^2+4 x_2 x_1+4 x_3 x_1+3 x_2 x_3\right)}{12 x_1^4}\ln x_2\, \ln x_3 + {(2x_2-x_3)(2x_3-x_2)\over 24x_2x_3}\ln x_1\\
&\,
+{i\pi\over 4}
-{x_2 Q_{23}(x_1,x_2,x_3)\over24(1-x_2)^2x_1^3x_3}\ln x_2
-{x_3 Q_{23}(x_1,x_3,x_2)\over24(1-x_3)^2x_1^3x_2}\ln x_3\,.
\esp\eeq

\beq\bsp
\overline{E}_\beta^{(2)} &\,= -{i\pi^3\over 48}-{i\pi\over3}\,A_\alpha^{(1)} 
+\frac{1}{18}\ln^3 x_1-\frac{1}{12} \ln^2 x_1 \ln (x_2x_3)-\frac{1}{12} \ln^2 x_2 \ln(x_1x_3)\\
&\,-\frac{1}{12} \ln^2 x_3 \ln(x_1x_2)+\frac{1}{3}\ln x_1\, \ln x_2\, \ln x_3+\frac{\pi ^2}{144}\ln x_1+\frac{1}{6} \ln(x_2x_3) \text{Li}_2\left(1-x_1\right)\\
&\,-\frac{Q_{14}(x_1,x_2,x_3)}{36 x_1^3}\ln^3 x_2+\frac{Q_{26}(x_1,x_3,x_2)}{36 x_1^3}\ln^3 x_3-\frac{1}{3} \text{Li}_3\left(1-\frac{1}{x_1}\right)-\frac{2}{3} \text{Li}_3\left(1-x_1\right)\\
&\,+\frac{1}{12} \text{Li}_2\left(1-x_2\right) \left[\frac{Q_{15}(x_1,x_2,x_3)}{x_1^3}\ln x_2+\frac{Q_{16}(x_1,x_2,x_3)}{x_1^3}\ln x_3\right]\\
&\,+\frac{1}{12} \text{Li}_2\left(1-x_3\right) \left[\frac{Q_{16}(x_1,x_3,x_2)}{x_1^3}\ln x_2+\frac{Q_{15}(x_1,x_3,x_2)}{x_1^3}\ln x_3\right]\\
&\,+\frac{\left(3 x_1+2 x_2\right) \left(3 x_1^2+2 x_2^2\right)}{144 x_1^3}\pi ^2 \ln x_2+\frac{\left(3 x_1+2 x_3\right) \left(3 x_1^2+2 x_3^2\right)}{144 x_1^3}\pi ^2 \ln x_3\\
&\,+\frac{Q_{14}(x_1,x_2,x_3)}{6 x_1^3}\text{Li}_3\left(1-\frac{1}{x_2}\right)+\frac{Q_{14}(x_1,x_3,x_2)}{6 x_1^3}\text{Li}_3\left(1-\frac{1}{x_3}\right)\\
&\,
+\frac{Q_{13}(x_1,x_2,x_3)}{6 x_1^3}\text{Li}_3\left(1-x_2\right)+\frac{Q_{13}(x_1,x_3,x_2)}{6 x_1^3}\text{Li}_3\left(1-x_3\right)\\
&\,+\frac{Q_{17}(x_1,x_2,x_3)}{24 x_1^3} \zeta_3-\frac{11}{36} \left(\ln^2 x_1+\ln^2 x_2+\ln^2 x_3\right)+{11\over18} i\pi \ln(x_1x_2x_3)\\
&\,-\frac{1}{36} \ln x_1\, \left[\frac{33 x_2^2+42 x_3 x_2-10 x_3^2}{x_3^2}\ln x_2+\frac{-10 x_2^2+42 x_3 x_2+33 x_3^2}{x_2^2}\ln x_3\right]\\
&\,+\frac{Q_{20}(x_1,x_2,x_3)}{36 x_1^4}\ln x_2\, \ln x_3-\frac{Q_{18}(x_1,x_2,x_3)}{36 x_2^2 x_3^2}\text{Li}_2\left(1-x_1\right)+\frac{Q_{19}(x_1,x_2,x_3)}{36 x_1^4 x_3^2}\text{Li}_2\left(1-x_2\right)\\
&\,+\frac{Q_{19}(x_1,x_3,x_2)}{36 x_1^4 x_2^2}\text{Li}_2\left(1-x_3\right)+\frac{ Q_{21}(x_1,x_2,x_3)}{288 x_1^4 x_2^2 x_3^2}\pi ^2 + \frac{Q_{12}\left(x_1,x_2,x_3\right)}{216 x_1^2 x_2 x_3}\ln x_1\\
&\,+\frac{Q_{11}\left(x_1,x_2,x_3\right)}{216 x_1^3 \left(1-x_2\right)^2 x_3}\ln x_2+\frac{Q_{11}\left(x_1,x_3,x_2\right)}{216 x_1^3 x_2 \left(1-x_3\right)^2}\ln x_3-\frac{142 x_1^2+
13 x_2 x_3}{36 x_1^2}i \pi\\
&\,  +\frac{x_2^2 x_3^2}{36 x_1^4}+\frac{135 x_1-748 x_2 x_3}{216 x_1^2}-\frac{13 \left(x_1+1\right)}{12 \left(1-x_2\right) \left(1-x_3\right)}-\frac{1115}{432}
\esp\eeq

\beq\bsp
\overline{F}_\beta^{(2)} &\,=
-{i\pi\over18}\ln(x_1x_2x_3) + {1\over36} \sum_{i=1}^3\ln^2x_i - {11\over144}\pi^2
-\frac{5 (x_2^2+x_3^2)+13 x_3 x_2-10 (x_2+x_3)}{54 x_1^2}\ln x_1\\
&\,
+\frac{5 (x_2^2+x_3^2)+11 x_3 x_2-10 (x_2+x_3)+5}{18 x_1^2}i \pi 
+\frac{x_2 x_3 }{216 x_1^4}\left(20 x_1^2-3 x_2 x_3\right)-\frac{5}{54} \ln (x_1x_2x_3)\,,
\esp\eeq
where the $Q_i(x,y,z)$ are homogeneous polynomials,
\beq\bsp
Q_{1}(x,y,z) &\,= -39 x^5 z+20 x^4 y^2+48 x^4 y z-78 x^4 z^2+30 x^3 \
y^2 z-20 x^3 y z^2-39 x^3 z^3\\
&\,-4 x^2 y^2 z^2-80 x^2 y z^3+78 x y^2 \
z^3-12 x y z^4+60 y^2 z^4\,,\\ 
Q_{2}(x,y,z) &\,= 60 x^2 \left(y^2+z^2\right)+9 x^2 y z-22 y^2 z^2\,,\
\\ 
Q_{3}(x,y,z) &\,= 33 x^3+12 x^2 y+12 x^2 z+3 x y^2+3 x z^2+2 y^3+2 \
z^3\,,\\ 
Q_{4}(x,y,z) &\,= 33 x^3+12 x^2 y+3 x y^2+2 y^3\,,\\ 
Q_{5}(x,y,z) &\,= -(x+z) \left(11 x^2+x z+2 z^2\right)\,,\\ 
Q_{6}(x,y,z) &\,= -99 x^3-48 x^2 y-12 x y^2-8 y^3\,,\\ 
Q_{7}(x,y,z) &\,= -561 x^3-96 x^2 y-96 x^2 z-24 x y^2-24 x z^2-16 \
y^3-16 z^3\,,\\ 
Q_{8}(x,y,z) &\,= 30 x^4 y^2+75 x^4 y z-134 x^4 z^2+6 x^3 y z^2-6 x^3 \
z^3+6 x^2 y^2 z^2-93 x^2 y z^3\\
&\,-6 x^2 z^4-24 x y^2 z^3-24 x y z^4+90 \
y^2 z^4\,,\\ 
Q_{9}(x,y,z) &\,= 30 y^4+75 y^3 z-134 y^2 z^2+75 y z^3+30 z^4\,,\\ 
Q_{10}(x,y,z) &\,= -20 x^4 y^4-50 x^4 y^3 z-115 x^4 y^2 z^2-50 x^4 y \
z^3-20 x^4 z^4-4 x^3 y^3 z^2\\
&\,-4 x^3 y^2 z^3-4 x^2 y^4 z^2+62 x^2 y^3 \
z^3-4 x^2 y^2 z^4+16 x y^4 z^3+16 x y^3 z^4-60 y^4 z^4\,,\\ 
Q_{11}(x,y,z) &\,= 206 x^5 z-198 x^4 y^2-234 x^4 y z+412 x^4 z^2-297 \
x^3 y^2 z+72 x^3 y z^2\\
&\,+206 x^3 z^3-126 x^2 y^2 z^2+342 x^2 y z^3-855 \
x y^2 z^3+36 x y z^4-594 y^2 z^4\,,\\ 
Q_{12}(x,y,z) &\,= -198 x^2 y^2+35 x^2 y z-198 x^2 z^2+78 y^2 z^2\,,\\ 
Q_{13}(x,y,z) &\,= -6 x^3-3 x^2 y-3 x^2 z-3 x y^2-3 x z^2-2 y^3-2 z^3\
\,,\\
Q_{14}(x,y,z) &\,= -6 x^3-3 x^2 y-3 x y^2-2 y^3\,,\\ 
Q_{15}(x,y,z) &\,= 4 x^3+3 x^2 y+3 x y^2+2 y^3\,,\\ 
Q_{16}(x,y,z) &\,= (x+z) \left(2 x^2+x z+2 z^2\right)\,,\\ 
Q_{17}(x,y,z) &\,= 27 x^3+12 x^2 y+12 x^2 z+12 x y^2+12 x z^2+8 y^3+8 \
z^3\,,\\ 
Q_{18}(x,y,z) &\,= 33 y^4+42 y^3 z-20 y^2 z^2+42 y z^3+33 z^4\,,\\ 
Q_{19}(x,y,z) &\,= -33 x^4 y^2-42 x^4 y z+20 x^4 z^2-6 x^3 y z^2+6 \
x^3 z^3-6 x^2 y^2 z^2+48 x^2 y z^3\\
&\,+6 x^2 z^4+12 x y^2 z^3+12 x y \
z^4-99 y^2 z^4\,,\\ 
Q_{20}(x,y,z) &\,= 10 x^4+48 x^2 y z+12 x y^2 z+12 x y z^2-99 y^2 z^2\
\,,\\
Q_{21}(x,y,z) &\,= 44 x^4 y^4+56 x^4 y^3 z+177 x^4 y^2 z^2+56 x^4 y \
z^3+44 x^4 z^4+8 x^3 y^3 z^2+8 x^3 y^2 z^3\\
&\,+8 x^2 y^4 z^2-64 x^2 y^3 \
z^3+8 x^2 y^2 z^4-16 x y^4 z^3-16 x y^3 z^4+132 y^4 z^4\,,\\ 
Q_{22}(x,y,z) &\,= 67 x^4+65 x^3 y+65 x^3 z-2 x^2 y^2+27 x^2 y z-2 \
x^2 z^2-26 x y^2 z-26 x y z^2\\
&\,-12 y^2 z^2\,,\\ 
Q_{23}(x,y,z) &\,= 2 x^4 y-4 x^4 z+3 x^3 y z+12 x^3 z^2+18 x^2 y \
z^2+24 x^2 z^3+17 x y z^3+8 x z^4\\
&\,+6 y z^4\,,\\ 
Q_{24}(x,y,z) &\,= x^2 y-2 x^2 z-2 x y z+4 x z^2+3 y z^2\,,\\ 
Q_{25}(x,y,z) &\,= x^4 y^4-2 x^4 y^3 z-2 x^4 y z^3+x^4 z^4+6 x^2 y^3 \
z^3+4 x y^4 z^3+4 x y^3 z^4+3 y^4 z^4\,.
\esp\eeq
As expected, the results involve only classical polylogarithms with arguments such that all functions are real in the region~\eqref{eq:+++_constraint}. Furthermore, the $\mathbb{Z}_2$ Bose symmetry is completely manifest. However, due to the reduced symmetry with respect to the first helicity configuration, the expressions are not as compact as in the previous case. It is worth noting that the weight four contribution is identical for both helicity configurations.

\subsection{Analytic continuation to the scattering region}
The expressions presented in the previous section are only valid in the decay region~\eqref{eq:+++_constraint}. In the rest of this section we show how to perform the analytic continuation to the scattering region. We have to distinguish the following cases,
\beq\bsp
\left.\begin{array}{rl}
g^+g^+&\,\to Hg^+\\
g^+g^+&\,\to Hg^-
\end{array}\right\}&\,\quad x_1>0 {\rm~~and~~} x_2,x_3<0\,,\\
g^+g^-\,\to Hg^+ \phantom{xx}&\,\quad x_2 >0 {\rm~~and~~} x_1, x_3<0\,.
\esp\eeq
In all cases the kinematic invariants are subject to the constraint
\beq
x_1+x_2+x_3=1\,,
\eeq
which simply expresses $s+t+u=m_H^2$.
In the following we only discuss the analytic continuation in the case where all gluons have a positive helicity, all other cases being similar.

In the decay region all invariants are positive and have a small positive imaginary part. The analytic continuation to the scattering region is then performed according to the prescription
\beq
s_{23} \to |s_{23}|\,e^{i\pi} {\rm~~and~~} s_{13} \to |s_{13}|\,e^{i\pi} \,,
\eeq
while all other invariants remain unchanged. This implies the following prescription for the dimensionless ratios,
\beq
x_1\to x_1 {\rm~~and~~} x_2 \to\overline{x}_2\,e^{i\pi} {\rm~~and~~} x_3 \to\overline{x}_3\,e^{i\pi}\,,
\eeq
where we defined $\overline{x}_i = |x_i| = -x_i$.

Using these prescriptions, the Kummer functions are analytically continued according to
\beq\bsp
\Lambda_n\left(-z\,e^{i\delta\pi}\right) &\,\to (n-1)!\sum_{k=0}^{n-1}{(-1)^{n-k}\over k!}\,
\left[\ln|z| + i\delta\pi \right]^k\,\li_{n-k}\left(-z\right)\,.
\esp\eeq
In addition, we need to analytically continue classical polylogarithms of the form
\beq\label{eq:li1-z}
\li_n\left(1-z\,e^{i\delta\pi}\right)\,,\qquad z > 0 {\rm~~and~~}\delta=\pm1\,.
\eeq
While the corresponding formulas could be obtained by the help of, \emph{e.g.}, the {\sc Mathematica} package {\sc HPL}~\cite{Maitre:2005uu}, we show how the analytic continuation formulas can be derived from the coproduct. Similar to the case of the inversion relations discussed in Section~\ref{sec:examples}, we proceed recursively in the weight. At weight one, we immediately obtain
\beq
\li_1\left(1-z\,e^{i\delta\pi}\right) = -\ln\left(z\,e^{i\delta\pi}\right) = -\ln z - i\delta\pi\,.
\eeq
At weight 2, we act with the coproduct, and drop all the $i\pi$ terms in all the factors of the coproduct except the first one,
\beq\bsp
\Delta_{1,1}\Bigg[\li_2\left(1-z\,e^{i\delta\pi}\right)\Bigg] &\,= 
\li_1\left(1-z\,e^{i\delta\pi}\right)\otimes \ln\left(1-z\,e^{i\delta\pi}\right)\\
&\,= -\ln z\otimes \ln(1+z) - i\pi\otimes\ln(1+z)\\
&\,=\Delta_{1,1}\Bigg[-\li_2(-z) -\ln z\ln(1+z) -i\pi\ln(1+z)\Bigg]\,.
\esp\eeq
Thus, we obtain
\beq
\li_2\left(1-z\,e^{i\delta\pi}\right) = -\li_2(-z) -\ln z\ln(1+z) -i\pi\ln(1+z) + c\,\pi^2\,,
\eeq
for some rational number $c$. Specializing to $z=0$ (where we are insensitive to the phase), we immediately obtain $c = {1\over6}$.

At weight 3, we first act with $\Delta_{1,1,1}$,
\beq\bsp
\Delta_{1,1,1}\Bigg[\li_3\left(1-z\,e^{i\delta\pi}\right)\Bigg] &\,= 
\li_1\left(1-z\,e^{i\delta\pi}\right)\otimes \ln\left(1-z\,e^{i\delta\pi}\right)\otimes \ln\left(1-z\,e^{i\delta\pi}\right)\\
&\,= 
-\ln z\otimes \ln\left(1+z\right)\otimes \ln\left(1+z\right)-i\delta\pi\otimes \ln\left(1+z\right)\otimes \ln\left(1+z\right)\\
&\,= \Delta_{1,1,1}\Bigg[\li_3\left({1\over1+z}\right) - {1\over6}\ln^3(1+z)- {i\over2}\delta\pi\ln^2(1 + z)\Bigg]\,.
\esp\eeq
In order to determine the terms proportional to $\pi^2$, we compute
\beq\bsp
\Delta_{2,1}&\Bigg[\li_3\left(1-z\,e^{i\delta\pi}\right)-\left(\li_3\left({1\over1+z}\right) - {1\over6}\ln^3(1+z)- {i\over2}\delta\pi\ln^2(1 + z)\right)\Bigg]\\
& = {1\over3}\pi^2\otimes\ln(1 + z) = \Delta_{2,1}\Bigg[{\pi^2\over3}\ln(1 + z)\Bigg]\,.
\esp\eeq
Thus,
\beq
\li_3\left(1-z\,e^{i\delta\pi}\right) = \li_3\left({1\over1+z}\right) - {1\over6}\ln^3(1+z)- {i\over2}\delta\pi\ln^2(1 + z) + {\pi^2\over3}\ln(1 + z) + c\,\zeta_3\,.
\eeq
Specializing to $z=0$, we obtain $c=0$.

Using this technique we can recursively derive all the analytic continuation formulas for functions of the type~\eqref{eq:li1-z}. In particular, at weight 4 we obtain
\beq
\li_4\left(1-z\,e^{i\delta\pi}\right) =
-\li_4\left({1\over1 + z}\right) - {1\over24} \ln^4(1 + z) -{i \delta\pi\over6} \ln^3(1 + z) + 
 {\pi^2\over6} \ln^2(1 + z) + {\pi^4\over45}\,.
 \eeq
These formulas are enough to perform the analytic continuation from the decay region to the scattering region. We checked numerically that our results agree (after analytic continuation) with the results in the scattering region presented in Ref.~\cite{Gehrmann:2011aa} for all the helicity configurations.

\section{Conclusion}
\label{sec:conclusion}
While recent advances seem to indicate that, at least in the context of the $\cN=4$ Super Yang-Mills theory, scattering amplitudes are simpler than expected, it is a known fact that the analytical evaluation of multi-loop Feynman integrals can lead to very lengthy and complicated expressions involving new classes of transcendental functions only poorly studied in the literature. 
A systematic approach to study these new functions and their functional equations is therefore highly desirable, not only from the formal standpoint, but also in perspective of phenomenological applications. A first step in this direction has been made in Ref.~\cite{Goncharov:2010jf} with the introduction of the so-called symbol map that allows to map the combinatorics of transcendental functions defined by iterated integrals to the combinatorics in a certain tensor algebra.

In this paper we have proposed a novel approach to deal with complicated expressions that can arise from a special class of Feynman integral computations, namely those that can be evaluated in terms of multiple polylogarithms. The cornerstone of this approach is the coproduct on multiple polylogarithms introduced by Goncharov in Ref.~\cite{Goncharov-Galois}, augmented by some ideas from a recent paper by Brown~\cite{Brown:2011ik}. The main feature is that, unlike the symbol, the coproduct allows to incorporate $\zeta$ values into the calculus, thus retaining more information about the function. We have demonstrated the virtue of this novel approach by rewriting the two-loop helicity amplitudes for a Higgs boson plus three gluons originally computed in Ref.~\cite{Koukoutsakis,Gehrmann:2011aa} in a compact analytical form, revealing, at least to our knowledge, for the first time an unexpected simplicity for a two-loop multi-scale amplitude in QCD.

\section*{Acknowledgements}
The author is grateful to Vittorio~Del~Duca, Herbert~Gangl and Volodya Smirnov for useful discussions and to Thomas Gehrmann and Nigel Glover for valuable comments on the manuscript. This work was supported by the ERC grant ``IterQCD''.


\appendix

\section{The multiple polylogarithm coproduct}
\label{app:coproduct}
\subsection{The coproduct in the generic case}
Before discussing how to define the coproduct of a multiple polylogarithm with non-generic arguments, let us first review a diagrammatic interpretation of the coproduct introduced in Refs.~\cite{Goncharov-Galois,Goncharov_Mixed_Tate}. 

{Consider a multiple polylogarithm of the form $I(a_0;a_1,\ldots,a_n;a_{n+1})$. We associate to it a diagram where we arrange the arguments of the multiple polylogarithm on a semi-circle. For example, for $n=3$ we have
\beq
\begin{fmffile}{semicircle}
\begin{fmfgraph*}(100,70)
\fmfbottom{a0,a5}
\fmfdot{a0}
\fmfdot{a5}
\fmfv{label=$a_0$}{a0}
\fmfv{label=$a_5$}{a5}
\fmf{plain}{a0,a5}
\fmffreeze
\fmf{plain,left,tension=0.2,tag=1}{a0,a5}
\fmfposition
\fmfipath{p[]}
\fmfiset{p1}{vpath1(__a0,__a5)}
\fmfiv{d.sh=circle,d.f=1,d.size=2thick,label=$a_1$}{point length(p1)/5 of p1}
\fmfiv{d.sh=circle,d.f=1,d.size=2thick,label=$a_2$}{point 2length(p1)/5 of p1}
\fmfiv{d.sh=circle,d.f=1,d.size=2thick,label=$a_3$}{point 3length(p1)/5 of p1}
\fmfiv{d.sh=circle,d.f=1,d.size=2thick,label=$a_4$}{point 4length(p1)/5 of p1}
\fmffreeze
\end{fmfgraph*}
\end{fmffile}
\eeq

The different terms in the coproduct in Eq.~\eqref{eq:coproduct} then correspond to connecting points via a polygon (including the empty polygon) in all possible ways. The points lying on the polygon provide the arguments for polylogarithm in the first factor in a given term in the coproduct, while the remaining points determine the entry of the second factor. Here we illustrate this construction only on the example of $I(a_0;a_1,a_2,a_3,a_4;a_{5})$, and we refer to Refs.~\cite{Goncharov-Galois,Goncharov_Mixed_Tate} for further details. In this case the polygons together with the terms in the coproduct they correspond to are\\
\begin{center}
\begin{tabular}{cc}
\begin{fmffile}{term1}
\begin{fmfgraph*}(100,70)
\fmfbottom{a0,a5}
\fmfdot{a0}
\fmfdot{a5}
\fmfv{label=$a_0$}{a0}
\fmfv{label=$a_5$}{a5}
\fmf{plain}{a0,a5}
\fmffreeze
\fmf{plain,left,tension=0.2,tag=1}{a0,a5}
\fmfposition
\fmfipath{p[]}
\fmfiset{p1}{vpath1(__a0,__a5)}
\fmfiv{d.sh=circle,d.f=1,d.size=2thick,label=$a_1$}{point length(p1)/5 of p1}
\fmfiv{d.sh=circle,d.f=1,d.size=2thick,label=$a_2$}{point 2length(p1)/5 of p1}
\fmfiv{d.sh=circle,d.f=1,d.size=2thick,label=$a_3$}{point 3length(p1)/5 of p1}
\fmfiv{d.sh=circle,d.f=1,d.size=2thick,label=$a_4$}{point 4length(p1)/5 of p1}
\fmffreeze
\end{fmfgraph*}
\end{fmffile}
&
\begin{fmffile}{term2}
\begin{fmfgraph*}(100,70)
\fmfbottom{a0,a5}
\fmfdot{a0}
\fmfdot{a5}
\fmfv{label=$a_0$}{a0}
\fmfv{label=$a_5$}{a5}
\fmf{plain}{a0,a5}
\fmffreeze
\fmf{plain,left,tension=0.2,tag=1}{a0,a5}
\fmfposition
\fmfipath{p[]}
\fmfiset{p1}{vpath1(__a0,__a5)}
\fmfiv{d.sh=circle,d.f=1,d.size=2thick,label=$a_1$}{point length(p1)/5 of p1}
\fmfiv{d.sh=circle,d.f=1,d.size=2thick,label=$a_2$}{point 2length(p1)/5 of p1}
\fmfiv{d.sh=circle,d.f=1,d.size=2thick,label=$a_3$}{point 3length(p1)/5 of p1}
\fmfiv{d.sh=circle,d.f=1,d.size=2thick,label=$a_4$}{point 4length(p1)/5 of p1}
\fmffreeze\fmfposition
\fmfipair{a[]}
\fmfiequ{a1}{point length(p1)/5 of p1}
\fmfiequ{a2}{point 2length(p1)/5 of p1}
\fmfiequ{a3}{point 3length(p1)/5 of p1}
\fmfiequ{a4}{point 4length(p1)/5 of p1}
\fmfi{plain}{vloc(__a0) -- a1}
\fmfi{plain}{a1 -- vloc(__a5)}
\end{fmfgraph*}
\end{fmffile}
 \\
${\displaystyle1\otimes I(a_0;a_1,a_2,a_3,a_4;a_5)}$ & ${\displaystyle I(a_0;a_1;a_5)\otimes I(a_1;a_2,a_3,a_4;a_5)}$ \\
$\phantom{aaa}$&\\
\begin{fmffile}{term3}
\begin{fmfgraph*}(100,70)
\fmfbottom{a0,a5}
\fmfdot{a0}
\fmfdot{a5}
\fmfv{label=$a_0$}{a0}
\fmfv{label=$a_5$}{a5}
\fmf{plain}{a0,a5}
\fmffreeze
\fmf{plain,left,tension=0.2,tag=1}{a0,a5}
\fmfposition
\fmfipath{p[]}
\fmfiset{p1}{vpath1(__a0,__a5)}
\fmfiv{d.sh=circle,d.f=1,d.size=2thick,label=$a_1$}{point length(p1)/5 of p1}
\fmfiv{d.sh=circle,d.f=1,d.size=2thick,label=$a_2$}{point 2length(p1)/5 of p1}
\fmfiv{d.sh=circle,d.f=1,d.size=2thick,label=$a_3$}{point 3length(p1)/5 of p1}
\fmfiv{d.sh=circle,d.f=1,d.size=2thick,label=$a_4$}{point 4length(p1)/5 of p1}
\fmffreeze\fmfposition
\fmfipair{a[]}
\fmfiequ{a1}{point length(p1)/5 of p1}
\fmfiequ{a2}{point 2length(p1)/5 of p1}
\fmfiequ{a3}{point 3length(p1)/5 of p1}
\fmfiequ{a4}{point 4length(p1)/5 of p1}
\fmfi{plain}{vloc(__a0) -- a2}
\fmfi{plain}{a2 -- vloc(__a5)}
\end{fmfgraph*}
\end{fmffile}
&
\begin{fmffile}{term4}
\begin{fmfgraph*}(100,70)
\fmfbottom{a0,a5}
\fmfdot{a0}
\fmfdot{a5}
\fmfv{label=$a_0$}{a0}
\fmfv{label=$a_5$}{a5}
\fmf{plain}{a0,a5}
\fmffreeze
\fmf{plain,left,tension=0.2,tag=1}{a0,a5}
\fmfposition
\fmfipath{p[]}
\fmfiset{p1}{vpath1(__a0,__a5)}
\fmfiv{d.sh=circle,d.f=1,d.size=2thick,label=$a_1$}{point length(p1)/5 of p1}
\fmfiv{d.sh=circle,d.f=1,d.size=2thick,label=$a_2$}{point 2length(p1)/5 of p1}
\fmfiv{d.sh=circle,d.f=1,d.size=2thick,label=$a_3$}{point 3length(p1)/5 of p1}
\fmfiv{d.sh=circle,d.f=1,d.size=2thick,label=$a_4$}{point 4length(p1)/5 of p1}
\fmffreeze\fmfposition
\fmfipair{a[]}
\fmfiequ{a1}{point length(p1)/5 of p1}
\fmfiequ{a2}{point 2length(p1)/5 of p1}
\fmfiequ{a3}{point 3length(p1)/5 of p1}
\fmfiequ{a4}{point 4length(p1)/5 of p1}
\fmfi{plain}{vloc(__a0) -- a3}
\fmfi{plain}{a3 -- vloc(__a5)}
\end{fmfgraph*}
\end{fmffile}
 \\
${\displaystyle I(a_0;a_2;a_5)\otimes[I(a_0;a_1;a_2)I(a_2;a_3,a_4;a_5)] }$&  ${\displaystyle I(a_0;a_3;a_5)\otimes[I(a_0;a_1,a_2;a_3)I(a_3;a_4;a_5)]}$\\
$\phantom{aaa}$&\\
\begin{fmffile}{term5}
\begin{fmfgraph*}(100,70)
\fmfbottom{a0,a5}
\fmfdot{a0}
\fmfdot{a5}
\fmfv{label=$a_0$}{a0}
\fmfv{label=$a_5$}{a5}
\fmf{plain}{a0,a5}
\fmffreeze
\fmf{plain,left,tension=0.2,tag=1}{a0,a5}
\fmfposition
\fmfipath{p[]}
\fmfiset{p1}{vpath1(__a0,__a5)}
\fmfiv{d.sh=circle,d.f=1,d.size=2thick,label=$a_1$}{point length(p1)/5 of p1}
\fmfiv{d.sh=circle,d.f=1,d.size=2thick,label=$a_2$}{point 2length(p1)/5 of p1}
\fmfiv{d.sh=circle,d.f=1,d.size=2thick,label=$a_3$}{point 3length(p1)/5 of p1}
\fmfiv{d.sh=circle,d.f=1,d.size=2thick,label=$a_4$}{point 4length(p1)/5 of p1}
\fmffreeze\fmfposition
\fmfipair{a[]}
\fmfiequ{a1}{point length(p1)/5 of p1}
\fmfiequ{a2}{point 2length(p1)/5 of p1}
\fmfiequ{a3}{point 3length(p1)/5 of p1}
\fmfiequ{a4}{point 4length(p1)/5 of p1}
\fmfi{plain}{vloc(__a0) -- a4}
\fmfi{plain}{a4 -- vloc(__a5)}
\end{fmfgraph*}
\end{fmffile}
&
\begin{fmffile}{term6}
\begin{fmfgraph*}(100,70)
\fmfbottom{a0,a5}
\fmfdot{a0}
\fmfdot{a5}
\fmfv{label=$a_0$}{a0}
\fmfv{label=$a_5$}{a5}
\fmf{plain}{a0,a5}
\fmffreeze
\fmf{plain,left,tension=0.2,tag=1}{a0,a5}
\fmfposition
\fmfipath{p[]}
\fmfiset{p1}{vpath1(__a0,__a5)}
\fmfiv{d.sh=circle,d.f=1,d.size=2thick,label=$a_1$}{point length(p1)/5 of p1}
\fmfiv{d.sh=circle,d.f=1,d.size=2thick,label=$a_2$}{point 2length(p1)/5 of p1}
\fmfiv{d.sh=circle,d.f=1,d.size=2thick,label=$a_3$}{point 3length(p1)/5 of p1}
\fmfiv{d.sh=circle,d.f=1,d.size=2thick,label=$a_4$}{point 4length(p1)/5 of p1}
\fmffreeze\fmfposition
\fmfipair{a[]}
\fmfiequ{a1}{point length(p1)/5 of p1}
\fmfiequ{a2}{point 2length(p1)/5 of p1}
\fmfiequ{a3}{point 3length(p1)/5 of p1}
\fmfiequ{a4}{point 4length(p1)/5 of p1}
\fmfi{plain}{vloc(__a0) -- a1}
\fmfi{plain}{a1-- a2}
\fmfi{plain}{a2 -- vloc(__a5)}
\end{fmfgraph*}
\end{fmffile}
 \\
${\displaystyle I(a_0;a_4;a_5)\otimes I(a_0;a_1,a_2,a_3;a_4) }$&  ${\displaystyle I(a_0;a_1,a_2;a_5)\otimes I(a_2;a_3,a_4;a_5)}$\\
$\phantom{aaa}$&\\
\begin{fmffile}{term7}
\begin{fmfgraph*}(100,70)
\fmfbottom{a0,a5}
\fmfdot{a0}
\fmfdot{a5}
\fmfv{label=$a_0$}{a0}
\fmfv{label=$a_5$}{a5}
\fmf{plain}{a0,a5}
\fmffreeze
\fmf{plain,left,tension=0.2,tag=1}{a0,a5}
\fmfposition
\fmfipath{p[]}
\fmfiset{p1}{vpath1(__a0,__a5)}
\fmfiv{d.sh=circle,d.f=1,d.size=2thick,label=$a_1$}{point length(p1)/5 of p1}
\fmfiv{d.sh=circle,d.f=1,d.size=2thick,label=$a_2$}{point 2length(p1)/5 of p1}
\fmfiv{d.sh=circle,d.f=1,d.size=2thick,label=$a_3$}{point 3length(p1)/5 of p1}
\fmfiv{d.sh=circle,d.f=1,d.size=2thick,label=$a_4$}{point 4length(p1)/5 of p1}
\fmffreeze\fmfposition
\fmfipair{a[]}
\fmfiequ{a1}{point length(p1)/5 of p1}
\fmfiequ{a2}{point 2length(p1)/5 of p1}
\fmfiequ{a3}{point 3length(p1)/5 of p1}
\fmfiequ{a4}{point 4length(p1)/5 of p1}
\fmfi{plain}{vloc(__a0) -- a1}
\fmfi{plain}{ a1 -- a3}
\fmfi{plain}{a3 -- vloc(__a5)}
\end{fmfgraph*}
\end{fmffile}
&
\begin{fmffile}{term8}
\begin{fmfgraph*}(100,70)
\fmfbottom{a0,a5}
\fmfdot{a0}
\fmfdot{a5}
\fmfv{label=$a_0$}{a0}
\fmfv{label=$a_5$}{a5}
\fmf{plain}{a0,a5}
\fmffreeze
\fmf{plain,left,tension=0.2,tag=1}{a0,a5}
\fmfposition
\fmfipath{p[]}
\fmfiset{p1}{vpath1(__a0,__a5)}
\fmfiv{d.sh=circle,d.f=1,d.size=2thick,label=$a_1$}{point length(p1)/5 of p1}
\fmfiv{d.sh=circle,d.f=1,d.size=2thick,label=$a_2$}{point 2length(p1)/5 of p1}
\fmfiv{d.sh=circle,d.f=1,d.size=2thick,label=$a_3$}{point 3length(p1)/5 of p1}
\fmfiv{d.sh=circle,d.f=1,d.size=2thick,label=$a_4$}{point 4length(p1)/5 of p1}
\fmffreeze\fmfposition
\fmfipair{a[]}
\fmfiequ{a1}{point length(p1)/5 of p1}
\fmfiequ{a2}{point 2length(p1)/5 of p1}
\fmfiequ{a3}{point 3length(p1)/5 of p1}
\fmfiequ{a4}{point 4length(p1)/5 of p1}
\fmfi{plain}{vloc(__a0) -- a1}
\fmfi{plain}{a1-- a4}
\fmfi{plain}{a4 -- vloc(__a5)}
\end{fmfgraph*}
\end{fmffile}
 \\
${\displaystyle I(a_0;a_1,a_3;a_5)\otimes [I(a_1;a_2;a_3)I(a_3;a_4;a_5)] }$&  ${\displaystyle I(a_0;a_1,a_4;a_5)\otimes I(a_1;a_2,a_3;a_4)}$

\end{tabular}
\end{center}
\begin{center}
\begin{tabular}{cc}
\begin{fmffile}{term9}
\begin{fmfgraph*}(100,70)
\fmfbottom{a0,a5}
\fmfdot{a0}
\fmfdot{a5}
\fmfv{label=$a_0$}{a0}
\fmfv{label=$a_5$}{a5}
\fmf{plain}{a0,a5}
\fmffreeze
\fmf{plain,left,tension=0.2,tag=1}{a0,a5}
\fmfposition
\fmfipath{p[]}
\fmfiset{p1}{vpath1(__a0,__a5)}
\fmfiv{d.sh=circle,d.f=1,d.size=2thick,label=$a_1$}{point length(p1)/5 of p1}
\fmfiv{d.sh=circle,d.f=1,d.size=2thick,label=$a_2$}{point 2length(p1)/5 of p1}
\fmfiv{d.sh=circle,d.f=1,d.size=2thick,label=$a_3$}{point 3length(p1)/5 of p1}
\fmfiv{d.sh=circle,d.f=1,d.size=2thick,label=$a_4$}{point 4length(p1)/5 of p1}
\fmffreeze\fmfposition
\fmfipair{a[]}
\fmfiequ{a1}{point length(p1)/5 of p1}
\fmfiequ{a2}{point 2length(p1)/5 of p1}
\fmfiequ{a3}{point 3length(p1)/5 of p1}
\fmfiequ{a4}{point 4length(p1)/5 of p1}
\fmfi{plain}{vloc(__a0) -- a2}
\fmfi{plain}{ a2 -- a3}
\fmfi{plain}{a3 -- vloc(__a5)}
\end{fmfgraph*}
\end{fmffile}
&
\begin{fmffile}{term10}
\begin{fmfgraph*}(100,70)
\fmfbottom{a0,a5}
\fmfdot{a0}
\fmfdot{a5}
\fmfv{label=$a_0$}{a0}
\fmfv{label=$a_5$}{a5}
\fmf{plain}{a0,a5}
\fmffreeze
\fmf{plain,left,tension=0.2,tag=1}{a0,a5}
\fmfposition
\fmfipath{p[]}
\fmfiset{p1}{vpath1(__a0,__a5)}
\fmfiv{d.sh=circle,d.f=1,d.size=2thick,label=$a_1$}{point length(p1)/5 of p1}
\fmfiv{d.sh=circle,d.f=1,d.size=2thick,label=$a_2$}{point 2length(p1)/5 of p1}
\fmfiv{d.sh=circle,d.f=1,d.size=2thick,label=$a_3$}{point 3length(p1)/5 of p1}
\fmfiv{d.sh=circle,d.f=1,d.size=2thick,label=$a_4$}{point 4length(p1)/5 of p1}
\fmffreeze\fmfposition
\fmfipair{a[]}
\fmfiequ{a1}{point length(p1)/5 of p1}
\fmfiequ{a2}{point 2length(p1)/5 of p1}
\fmfiequ{a3}{point 3length(p1)/5 of p1}
\fmfiequ{a4}{point 4length(p1)/5 of p1}
\fmfi{plain}{vloc(__a0) -- a2}
\fmfi{plain}{a2-- a4}
\fmfi{plain}{a4 -- vloc(__a5)}
\end{fmfgraph*}
\end{fmffile}
 \\
${\displaystyle I(a_0;a_2,a_3;a_5)\otimes [I(a_0;a_1;a_2)I(a_3;a_4;a_5)] }$&  ${\displaystyle I(a_0;a_2,a_4;a_5)\otimes [I(a_0;a_1;a_2)I(a_2;a_3;a_4)]}$\\
$\phantom{aaa}$&\\
\begin{fmffile}{term11}
\begin{fmfgraph*}(100,70)
\fmfbottom{a0,a5}
\fmfdot{a0}
\fmfdot{a5}
\fmfv{label=$a_0$}{a0}
\fmfv{label=$a_5$}{a5}
\fmf{plain}{a0,a5}
\fmffreeze
\fmf{plain,left,tension=0.2,tag=1}{a0,a5}
\fmfposition
\fmfipath{p[]}
\fmfiset{p1}{vpath1(__a0,__a5)}
\fmfiv{d.sh=circle,d.f=1,d.size=2thick,label=$a_1$}{point length(p1)/5 of p1}
\fmfiv{d.sh=circle,d.f=1,d.size=2thick,label=$a_2$}{point 2length(p1)/5 of p1}
\fmfiv{d.sh=circle,d.f=1,d.size=2thick,label=$a_3$}{point 3length(p1)/5 of p1}
\fmfiv{d.sh=circle,d.f=1,d.size=2thick,label=$a_4$}{point 4length(p1)/5 of p1}
\fmffreeze\fmfposition
\fmfipair{a[]}
\fmfiequ{a1}{point length(p1)/5 of p1}
\fmfiequ{a2}{point 2length(p1)/5 of p1}
\fmfiequ{a3}{point 3length(p1)/5 of p1}
\fmfiequ{a4}{point 4length(p1)/5 of p1}
\fmfi{plain}{vloc(__a0) -- a3}
\fmfi{plain}{ a3 -- a4}
\fmfi{plain}{a4 -- vloc(__a5)}
\end{fmfgraph*}
\end{fmffile}
&
\begin{fmffile}{term12}
\begin{fmfgraph*}(100,70)
\fmfbottom{a0,a5}
\fmfdot{a0}
\fmfdot{a5}
\fmfv{label=$a_0$}{a0}
\fmfv{label=$a_5$}{a5}
\fmf{plain}{a0,a5}
\fmffreeze
\fmf{plain,left,tension=0.2,tag=1}{a0,a5}
\fmfposition
\fmfipath{p[]}
\fmfiset{p1}{vpath1(__a0,__a5)}
\fmfiv{d.sh=circle,d.f=1,d.size=2thick,label=$a_1$}{point length(p1)/5 of p1}
\fmfiv{d.sh=circle,d.f=1,d.size=2thick,label=$a_2$}{point 2length(p1)/5 of p1}
\fmfiv{d.sh=circle,d.f=1,d.size=2thick,label=$a_3$}{point 3length(p1)/5 of p1}
\fmfiv{d.sh=circle,d.f=1,d.size=2thick,label=$a_4$}{point 4length(p1)/5 of p1}
\fmffreeze\fmfposition
\fmfipair{a[]}
\fmfiequ{a1}{point length(p1)/5 of p1}
\fmfiequ{a2}{point 2length(p1)/5 of p1}
\fmfiequ{a3}{point 3length(p1)/5 of p1}
\fmfiequ{a4}{point 4length(p1)/5 of p1}
\fmfi{plain}{vloc(__a0) -- a1}
\fmfi{plain}{a1-- a2}
\fmfi{plain}{a2-- a3}
\fmfi{plain}{a3 -- vloc(__a5)}
\end{fmfgraph*}
\end{fmffile}
 \\
${\displaystyle I(a_0;a_3,a_4;a_5)\otimes I(a_0;a_1,a_2;a_3) }$&  ${\displaystyle I(a_0;a_1,a_2,a_3;a_5)\otimes  I(a_3;a_4;a_5)}$\\
$\phantom{aaa}$&\\
\begin{fmffile}{term13}
\begin{fmfgraph*}(100,70)
\fmfbottom{a0,a5}
\fmfdot{a0}
\fmfdot{a5}
\fmfv{label=$a_0$}{a0}
\fmfv{label=$a_5$}{a5}
\fmf{plain}{a0,a5}
\fmffreeze
\fmf{plain,left,tension=0.2,tag=1}{a0,a5}
\fmfposition
\fmfipath{p[]}
\fmfiset{p1}{vpath1(__a0,__a5)}
\fmfiv{d.sh=circle,d.f=1,d.size=2thick,label=$a_1$}{point length(p1)/5 of p1}
\fmfiv{d.sh=circle,d.f=1,d.size=2thick,label=$a_2$}{point 2length(p1)/5 of p1}
\fmfiv{d.sh=circle,d.f=1,d.size=2thick,label=$a_3$}{point 3length(p1)/5 of p1}
\fmfiv{d.sh=circle,d.f=1,d.size=2thick,label=$a_4$}{point 4length(p1)/5 of p1}
\fmffreeze\fmfposition
\fmfipair{a[]}
\fmfiequ{a1}{point length(p1)/5 of p1}
\fmfiequ{a2}{point 2length(p1)/5 of p1}
\fmfiequ{a3}{point 3length(p1)/5 of p1}
\fmfiequ{a4}{point 4length(p1)/5 of p1}
\fmfi{plain}{vloc(__a0) -- a1}
\fmfi{plain}{ a1 -- a2}
\fmfi{plain}{ a2 -- a4}
\fmfi{plain}{a4 -- vloc(__a5)}
\end{fmfgraph*}
\end{fmffile}
&
\begin{fmffile}{term14}
\begin{fmfgraph*}(100,70)
\fmfbottom{a0,a5}
\fmfdot{a0}
\fmfdot{a5}
\fmfv{label=$a_0$}{a0}
\fmfv{label=$a_5$}{a5}
\fmf{plain}{a0,a5}
\fmffreeze
\fmf{plain,left,tension=0.2,tag=1}{a0,a5}
\fmfposition
\fmfipath{p[]}
\fmfiset{p1}{vpath1(__a0,__a5)}
\fmfiv{d.sh=circle,d.f=1,d.size=2thick,label=$a_1$}{point length(p1)/5 of p1}
\fmfiv{d.sh=circle,d.f=1,d.size=2thick,label=$a_2$}{point 2length(p1)/5 of p1}
\fmfiv{d.sh=circle,d.f=1,d.size=2thick,label=$a_3$}{point 3length(p1)/5 of p1}
\fmfiv{d.sh=circle,d.f=1,d.size=2thick,label=$a_4$}{point 4length(p1)/5 of p1}
\fmffreeze\fmfposition
\fmfipair{a[]}
\fmfiequ{a1}{point length(p1)/5 of p1}
\fmfiequ{a2}{point 2length(p1)/5 of p1}
\fmfiequ{a3}{point 3length(p1)/5 of p1}
\fmfiequ{a4}{point 4length(p1)/5 of p1}
\fmfi{plain}{vloc(__a0) -- a1}
\fmfi{plain}{a1-- a3}
\fmfi{plain}{a3-- a4}
\fmfi{plain}{a4 -- vloc(__a5)}
\end{fmfgraph*}
\end{fmffile}
 \\
${\displaystyle I(a_0;a_1,a_2,a_4;a_5)\otimes I(a_2;a_3;a_4) }$&  ${\displaystyle I(a_0;a_1,a_3,a_4;a_5)\otimes  I(a_1;a_2;a_3)}$\\
$\phantom{aaa}$&\\
\begin{fmffile}{term15}
\begin{fmfgraph*}(100,70)
\fmfbottom{a0,a5}
\fmfdot{a0}
\fmfdot{a5}
\fmfv{label=$a_0$}{a0}
\fmfv{label=$a_5$}{a5}
\fmf{plain}{a0,a5}
\fmffreeze
\fmf{plain,left,tension=0.2,tag=1}{a0,a5}
\fmfposition
\fmfipath{p[]}
\fmfiset{p1}{vpath1(__a0,__a5)}
\fmfiv{d.sh=circle,d.f=1,d.size=2thick,label=$a_1$}{point length(p1)/5 of p1}
\fmfiv{d.sh=circle,d.f=1,d.size=2thick,label=$a_2$}{point 2length(p1)/5 of p1}
\fmfiv{d.sh=circle,d.f=1,d.size=2thick,label=$a_3$}{point 3length(p1)/5 of p1}
\fmfiv{d.sh=circle,d.f=1,d.size=2thick,label=$a_4$}{point 4length(p1)/5 of p1}
\fmffreeze\fmfposition
\fmfipair{a[]}
\fmfiequ{a1}{point length(p1)/5 of p1}
\fmfiequ{a2}{point 2length(p1)/5 of p1}
\fmfiequ{a3}{point 3length(p1)/5 of p1}
\fmfiequ{a4}{point 4length(p1)/5 of p1}
\fmfi{plain}{vloc(__a0) -- a2}
\fmfi{plain}{a2-- a3}
\fmfi{plain}{a3-- a4}
\fmfi{plain}{a4 -- vloc(__a5)}
\end{fmfgraph*}
\end{fmffile}
&
\begin{fmffile}{term16}
\begin{fmfgraph*}(100,70)
\fmfbottom{a0,a5}
\fmfdot{a0}
\fmfdot{a5}
\fmfv{label=$a_0$}{a0}
\fmfv{label=$a_5$}{a5}
\fmf{plain}{a0,a5}
\fmffreeze
\fmf{plain,left,tension=0.2,tag=1}{a0,a5}
\fmfposition
\fmfipath{p[]}
\fmfiset{p1}{vpath1(__a0,__a5)}
\fmfiv{d.sh=circle,d.f=1,d.size=2thick,label=$a_1$}{point length(p1)/5 of p1}
\fmfiv{d.sh=circle,d.f=1,d.size=2thick,label=$a_2$}{point 2length(p1)/5 of p1}
\fmfiv{d.sh=circle,d.f=1,d.size=2thick,label=$a_3$}{point 3length(p1)/5 of p1}
\fmfiv{d.sh=circle,d.f=1,d.size=2thick,label=$a_4$}{point 4length(p1)/5 of p1}
\fmffreeze\fmfposition
\fmfipair{a[]}
\fmfiequ{a1}{point length(p1)/5 of p1}
\fmfiequ{a2}{point 2length(p1)/5 of p1}
\fmfiequ{a3}{point 3length(p1)/5 of p1}
\fmfiequ{a4}{point 4length(p1)/5 of p1}
\fmfi{plain}{vloc(__a0) -- a1}
\fmfi{plain}{a1-- a2}
\fmfi{plain}{a2-- a3}
\fmfi{plain}{a3-- a4}
\fmfi{plain}{a4 -- vloc(__a5)}
\end{fmfgraph*}
\end{fmffile}
 \\
${\displaystyle I(a_0;a_2,a_3,a_4;a_5)\otimes I(a_0;a_1;a_2) }$&  ${\displaystyle I(a_0;a_1,a_2,a_3,a_4;a_5)\otimes  1}$\\
\end{tabular}
\end{center}

\subsection{The coproduct in the non-generic case}
As already mentioned, the formula for the coproduct, Eq.~\eqref{eq:coproduct}, is only valid in the generic case where all the arguments are mutually different. Indeed, in the non-generic case divergences can arise in individual terms in the coproduct. A multiple polylogarithm $I(a_0;a_1,\ldots,a_n;a_{n+1})$ is in general divergent if either $a_{1} = a_0$ or $a_n=a_{n+1}$. In this case the poles in the integrand coincide with the endpoints of the integration path. As an example of how these divergences can arise inside the coproduct, consider the multiple polylogarithm $I(a_0;a_1,a_2,a_2;a_3)$, which is convergent whenever $a_0\neq a_1$ and $a_2\neq a_3$.
Eq.~\eqref{eq:coproduct}, however, contains a term
\begin{center}
\begin{tabular}{c}
\begin{fmffile}{divergence}
\begin{fmfgraph*}(100,70)
\fmfbottom{a0,a5}
\fmfdot{a0}
\fmfdot{a5}
\fmfv{label=$a_0$}{a0}
\fmfv{label=$a_3$}{a5}
\fmf{plain}{a0,a5}
\fmffreeze
\fmf{plain,left,tension=0.2,tag=1}{a0,a5}
\fmfposition
\fmfipath{p[]}
\fmfiset{p1}{vpath1(__a0,__a5)}
\fmfiv{d.sh=circle,d.f=1,d.size=2thick,label=$a_1$}{point length(p1)/4 of p1}
\fmfiv{d.sh=circle,d.f=1,d.size=2thick,label=$a_2$}{point 2length(p1)/4 of p1}
\fmfiv{d.sh=circle,d.f=1,d.size=2thick,label=$a_2$}{point 3length(p1)/4 of p1}
\fmffreeze\fmfposition
\fmfipair{a[]}
\fmfiequ{a1}{point length(p1)/4 of p1}
\fmfiequ{a2}{point 2length(p1)/4 of p1}
\fmfiequ{a3}{point 3length(p1)/4 of p1}
\fmfi{plain}{vloc(__a0) -- a1}
\fmfi{plain}{a1-- a3}
\fmfi{plain}{a3 -- vloc(__a5)}
\end{fmfgraph*}
\end{fmffile} \\
${\displaystyle I(a_0;a_1,a_2;a_3)\otimes I(a_1;a_2;a_2)\,,}$
\end{tabular}
\end{center}
and the second factor, $I(a_1;a_2;a_2)$, is divergent.

In Refs.~\cite{Goncharov-Galois,Goncharov_Mixed_Tate} the coproduct in the non-generic case is defined by replacing in the right-hand side of Eq.~\eqref{eq:coproduct} every multiple polylogarithm by its \emph{regularized value}. In the following we only give a practical rule of how to obtain a regularized value (there is more than one way to perform the regularization), and we refer to Refs.~\cite{Goncharov-Galois,Goncharov_Mixed_Tate} for more details.

As the divergences of a multiple polylogarithm are end-point divergences, we can easily regularize them by slightly moving the end points of the integration path. In practice we usually deal with integration paths that are straight lines, and in this case the regularization can simply be achieved by the replacement
\beq\label{eq:reg1}
I(a_0;a_1,\ldots,a_n;a_{n+1}) \to \left\{\begin{array}{ll} 
I(a_0(1+\varepsilon);a_1,\ldots,a_n;a_{n+1}(1-\varepsilon))\,, & \textrm{ if } a_0\neq 0\,,\\
I(\varepsilon;a_1,\ldots,a_n;a_{n+1}(1-\varepsilon))\,, & \textrm{ if } a_0= 0\,.
\end{array}\right.
\eeq
This regularization procedure has two obvious features:
\begin{enumerate}
\item If $I(a_0;a_1,\ldots,a_n;a_{n+1})$ is convergent, then 
\beq
\lim_{\varepsilon\to0} I(a_0(1+\varepsilon);a_1,\ldots,a_n;a_{n+1}(1-\varepsilon)) = I(a_0;a_1,\ldots,a_n;a_{n+1})\,.
\eeq
\item The regularization~\eqref{eq:reg1} preserves shuffle identities.
\end{enumerate}
As a consequence, we can use shuffle identities to extract all the divergences of the multiple polylogarithms, \emph{e.g.},
\beq\bsp\label{eq:regexample}
I(a_0&(1+\varepsilon);a_1,a_2,a_2;a_{2}(1-\varepsilon)) = I(a_0(1+\varepsilon);a_2,a_2;a_{2}(1-\varepsilon)) I(a_0(1+\varepsilon);a_1;a_{2}(1-\varepsilon)) \\
&\,\quad- I(a_0(1+\varepsilon);a_2,a_1,a_2;a_{2}(1-\varepsilon)) - I(a_0(1+\varepsilon);a_2,a_2,a_1;a_{2}(1-\varepsilon))\\
&\,=I(a_0(1+\varepsilon);a_2,a_2;a_{2}(1-\varepsilon)) I(a_0(1+\varepsilon);a_1;a_{2}(1-\varepsilon))\\
&\,
\quad- I(a_0(1+\varepsilon);a_2,a_1;a_{2}(1-\varepsilon))I(a_0(1+\varepsilon);a_2;a_{2}(1-\varepsilon)) \\
&\,\quad+I(a_0(1+\varepsilon);a_2,a_2;a_{1}(1-\varepsilon))\\
&\,={1\over 2}\ln^2\varepsilon \,I(a_0;a_1;a_{2}) - \ln\varepsilon\, I(a_0;a_2,a_1;a_{2}) + I(a_0;a_2,a_2,a_1;a_{2})\,.
\esp\eeq
In this way all the divergences are regularized, and the shuffle identities allow us to write the regularized integrals as a polynomials in $\ln\varepsilon$. The \emph{regularized value} $\hat{I}(a_0;a_1,\ldots,a_n;a_{n+1})$ of the multiple polylogarithm $I(a_0;a_1,\ldots,a_n;a_{n+1})$ is defined as the constant term of this polynomial. As an example, from Eq.~\eqref{eq:regexample} we obtain
\beq
\hat{I}(a_0;a_1,a_2,a_2;a_{2}) \equiv I(a_0;a_2,a_2,a_1;a_{2})\,.
\eeq
The coproduct in the non-generic case is then obtained by replacing all multiple polylogarithms in the right-hand side of Eq.~\eqref{eq:coproduct} by their regularized values,
 \beq\bsp
\Delta&(I(a_0;a_1,\ldots,a_n;a_{n+1})) \\
&= \sum_{0=i_1<i_2<\ldots<i_{k}<i_{k+1}=n} \hat{I}(a_0;a_{i_1},\ldots,a_{i_k};a_{n+1})\otimes\Bigg[\prod_{p=0}^k\hat{I}(a_{i_p};a_{i_p+1},\ldots,a_{i_{p+1}-1};a_{i_{p+1}})\Bigg]\,.
\esp\eeq

\section{Proofs of the derivative and monodromy identities}
\label{app:proofs}
\subsection{Proof of the derivative identity}
In this section we sketch the proof of the derivative identity~\eqref{eq:diff_mono} in a particular case, namely
\beq\label{eq:claim1}
\left(\id\otimes{\partial\over\partial a_{n+1}}\right)\Delta(I(a_0;a_1,\ldots,a_n;a_{n+1}))
=
\Delta\left({\partial\over\partial a_{n+1}}I(a_0;a_1,\ldots,a_n;a_{n+1})\right)\,,
\eeq
where we assume all arguments of the multiple polylogarithm generic. 

Let us compute the action of the differential operator in the left-hand side of Eq.~\eqref{eq:claim1}. It is obvious that we do only get a non-zero contribution from those terms in the coproduct where the second factor depends on $a_{n+1}$. If all the arguments are generic, it is easy to see that this is the case precisely for those terms in Eq.~\eqref{eq:coproduct} where the polygon inscribed into the semi-circle (see Appendix~\ref{app:coproduct}) does not contain $a_n$. As an example, for $n=3$, the term
\begin{center}
\begin{tabular}{c}
\begin{fmffile}{diff1}
\begin{fmfgraph*}(100,70)
\fmfbottom{a0,a5}
\fmfdot{a0}
\fmfdot{a5}
\fmfv{label=$a_0$}{a0}
\fmfv{label=$a_4$}{a5}
\fmf{plain}{a0,a5}
\fmffreeze
\fmf{plain,left,tension=0.2,tag=1}{a0,a5}
\fmfposition
\fmfipath{p[]}
\fmfiset{p1}{vpath1(__a0,__a5)}
\fmfiv{d.sh=circle,d.f=1,d.size=2thick,label=$a_1$}{point length(p1)/4 of p1}
\fmfiv{d.sh=circle,d.f=1,d.size=2thick,label=$a_2$}{point 2length(p1)/4 of p1}
\fmfiv{d.sh=circle,d.f=1,d.size=2thick,label=$a_3$}{point 3length(p1)/4 of p1}
\fmffreeze\fmfposition
\fmfipair{a[]}
\fmfiequ{a1}{point length(p1)/4 of p1}
\fmfiequ{a2}{point 2length(p1)/4 of p1}
\fmfiequ{a3}{point 3length(p1)/4 of p1}
\fmfi{plain}{vloc(__a0) -- a1}
\fmfi{plain}{a1-- a3}
\fmfi{plain}{a3 -- vloc(__a5)}
\end{fmfgraph*}
\end{fmffile} \\
${\displaystyle I(a_0;a_1,a_3;a_4)\otimes I(a_1;a_2;a_3)\,,}$
\end{tabular}
\end{center}
vanishes under the action of the differential operator $\left(\id\otimes{\partial\over\partial a_{n+1}}\right)$, whereas the term
\begin{center}
\begin{tabular}{c}
\begin{fmffile}{diff2}
\begin{fmfgraph*}(100,70)
\fmfbottom{a0,a5}
\fmfdot{a0}
\fmfdot{a5}
\fmfv{label=$a_0$}{a0}
\fmfv{label=$a_4$}{a5}
\fmf{plain}{a0,a5}
\fmffreeze
\fmf{plain,left,tension=0.2,tag=1}{a0,a5}
\fmfposition
\fmfipath{p[]}
\fmfiset{p1}{vpath1(__a0,__a5)}
\fmfiv{d.sh=circle,d.f=1,d.size=2thick,label=$a_1$}{point length(p1)/4 of p1}
\fmfiv{d.sh=circle,d.f=1,d.size=2thick,label=$a_2$}{point 2length(p1)/4 of p1}
\fmfiv{d.sh=circle,d.f=1,d.size=2thick,label=$a_3$}{point 3length(p1)/4 of p1}
\fmffreeze\fmfposition
\fmfipair{a[]}
\fmfiequ{a1}{point length(p1)/4 of p1}
\fmfiequ{a2}{point 2length(p1)/4 of p1}
\fmfiequ{a3}{point 3length(p1)/4 of p1}
\fmfi{plain}{vloc(__a0) -- a1}
\fmfi{plain}{a1 -- vloc(__a5)}
\end{fmfgraph*}
\end{fmffile} \\
${\displaystyle I(a_0;a_1;a_4)\otimes I(a_1;a_2,a_3;a_4)\,,}$
\end{tabular}
\end{center}
gives a non-zero contribution, namely
\beq
{1\over a_{4}-a_3}\times\left\{\begin{array}{c}
\begin{fmffile}{diff3}
\begin{fmfgraph*}(100,70)
\fmfbottom{a0,a5}
\fmfdot{a0}
\fmfdot{a5}
\fmfv{label=$a_0$}{a0}
\fmfv{label=$a_4$}{a5}
\fmf{plain}{a0,a5}
\fmffreeze
\fmf{plain,left,tension=0.2,tag=1}{a0,a5}
\fmfposition
\fmfipath{p[]}
\fmfiset{p1}{vpath1(__a0,__a5)}
\fmfiv{d.sh=circle,d.f=1,d.size=2thick,label=$a_1$}{point length(p1)/3 of p1}
\fmfiv{d.sh=circle,d.f=1,d.size=2thick,label=$a_2$}{point 2length(p1)/3 of p1}
\fmffreeze\fmfposition
\fmfipair{a[]}
\fmfiequ{a1}{point length(p1)/3 of p1}
\fmfiequ{a2}{point 2length(p1)/3 of p1}
\fmfi{plain}{vloc(__a0) -- a1}
\fmfi{plain}{a1 -- vloc(__a5)}
\end{fmfgraph*}
\end{fmffile} \\
I(a_0;a_1;a_4)\otimes I(a_1;a_2;a_4)\,,
\end{array}\right.
\eeq
It is clear that in this way we produce precisely all the polygons inscribed into the semi-circle with the point $a_n$ ($a_3$ in the example above) removed, multiplied by ${1\over a_{n+1}-a_n}$. These terms are in one-to-one correspondence with the terms in the coproduct of $I(a_0;a_1,\ldots,a_{n-1};a_{n+1})$. Thus, we obtain
\beq\bsp
\left(\id\otimes{\partial\over\partial a_{n+1}}\right)&\Delta(I(a_0;a_1,\ldots,a_n;a_{n+1}))
=
{1\over a_{n+1}-a_n}\,\Delta\left(I(a_0;a_1,\ldots,a_{n-1};a_{n+1})\right)\\
&\,=
\Delta\left({1\over a_{n+1}-a_n}\,I(a_0;a_1,\ldots,a_{n-1};a_{n+1})\right)\\
&\,=
\Delta\left({\partial\over\partial a_{n+1}}I(a_0;a_1,\ldots,a_{n};a_{n+1})\right)\,,
\esp\eeq
which finishes the proof.

\subsection{Proof of the monodromy identity}
In this section we sketch the proof of the monodromy identity~\eqref{eq:diff_mono} in a particular case, namely
\beq\label{eq:claim2}
\left(\cM_{a_{n+1}=a_i}\otimes\id\right)\Delta(I(a_0;a_1,\ldots,a_n;a_{n+1}))
=
\Delta\left(\cM_{a_{n+1}=a_i}I(a_0;a_1,\ldots,a_n;a_{n+1})\right)\,,
\eeq
 where all arguments of the multiple polylogarithm are assumed generic. 

We again start by evaluating the action of the monodromy operator in the left-hand side of Eq.~\eqref{eq:claim2}. It is clear that only those terms in the coproduct where the first factor depends on $a_i$ contribute a non-zero answer. For generic arguments this implies that the polygon associated to this term must contain $a_i$. As an example, for $n=3$ and $i=2$, the term
\begin{center}
\begin{tabular}{c}
\begin{fmffile}{mono1}
\begin{fmfgraph*}(100,70)
\fmfbottom{a0,a5}
\fmfdot{a0}
\fmfdot{a5}
\fmfv{label=$a_0$}{a0}
\fmfv{label=$a_4$}{a5}
\fmf{plain}{a0,a5}
\fmffreeze
\fmf{plain,left,tension=0.2,tag=1}{a0,a5}
\fmfposition
\fmfipath{p[]}
\fmfiset{p1}{vpath1(__a0,__a5)}
\fmfiv{d.sh=circle,d.f=1,d.size=2thick,label=$a_1$}{point length(p1)/4 of p1}
\fmfiv{d.sh=circle,d.f=1,d.size=2thick,label=$a_2$}{point 2length(p1)/4 of p1}
\fmfiv{d.sh=circle,d.f=1,d.size=2thick,label=$a_3$}{point 3length(p1)/4 of p1}
\fmffreeze\fmfposition
\fmfipair{a[]}
\fmfiequ{a1}{point length(p1)/4 of p1}
\fmfiequ{a2}{point 2length(p1)/4 of p1}
\fmfiequ{a3}{point 3length(p1)/4 of p1}
\fmfi{plain}{vloc(__a0) -- a1}
\fmfi{plain}{a1-- a3}
\fmfi{plain}{a3 -- vloc(__a5)}
\end{fmfgraph*}
\end{fmffile} \\
${\displaystyle I(a_0;a_1,a_3;a_4)\otimes I(a_1;a_2;a_3)\,,}$
\end{tabular}
\end{center}
vanishes under the action of the monodromy operator $\left(\cM_{a_{n+1}=a_i}\otimes\id\right)$, whereas the term
\begin{center}
\begin{tabular}{c}
\begin{fmffile}{mono2}
\begin{fmfgraph*}(100,70)
\fmfbottom{a0,a5}
\fmfdot{a0}
\fmfdot{a5}
\fmfv{label=$a_0$}{a0}
\fmfv{label=$a_4$}{a5}
\fmf{plain}{a0,a5}
\fmffreeze
\fmf{plain,left,tension=0.2,tag=1}{a0,a5}
\fmfposition
\fmfipath{p[]}
\fmfiset{p1}{vpath1(__a0,__a5)}
\fmfiv{d.sh=circle,d.f=1,d.size=2thick,label=$a_1$}{point length(p1)/4 of p1}
\fmfiv{d.sh=circle,d.f=1,d.size=2thick,label=$a_2$}{point 2length(p1)/4 of p1}
\fmfiv{d.sh=circle,d.f=1,d.size=2thick,label=$a_3$}{point 3length(p1)/4 of p1}
\fmffreeze\fmfposition
\fmfipair{a[]}
\fmfiequ{a1}{point length(p1)/4 of p1}
\fmfiequ{a2}{point 2length(p1)/4 of p1}
\fmfiequ{a3}{point 3length(p1)/4 of p1}
\fmfi{plain}{vloc(__a0) -- a1}
\fmfi{plain}{a1 -- a2}
\fmfi{plain}{a2 -- vloc(__a5)}
\end{fmfgraph*}
\end{fmffile} \\
${\displaystyle I(a_0;a_1,a_2;a_4)\otimes I(a_2;a_3;a_4)\,,}$
\end{tabular}
\end{center}
contributes non-trivially to the left-hand side of Eq.~\eqref{eq:claim2}. In Ref.~\cite{Goncharov_Mixed_Tate} a formula for the monodromy of a multiple polylogarithm was given,
\beq\label{eq:Gonch_mono}
\cM_{a_{n+1}=a_i}I(a_0;a_1,\ldots,a_n;a_{n+1}) = 2\pi i\,I(a_0;a_1,\ldots,a_{i-1};a_{i})\,I(a_i;a_{i+1},\ldots,a_n;a_{n+1})\,.
\eeq
Applying this formula to the left-hand side of Eq.~\eqref{eq:claim2} implies that the monodromy operator `splits' all the polygons that give a non-vanishing contribution at the point $a_i$. In the example $n=3$, $i=2$ considered above we obtain the splitting
\begin{center}
\begin{tabular}{c}
\begin{fmffile}{mono3}
\begin{fmfgraph*}(100,70)
\fmfbottom{a0,a5}
\fmfdot{a0}
\fmfdot{a5}
\fmfv{label=$a_0$}{a0}
\fmfv{label=$a_4$}{a5}
\fmf{plain}{a0,a5}
\fmffreeze
\fmf{plain,left,tension=0.2,tag=1}{a0,a5}
\fmfposition
\fmfipath{p[]}
\fmfiset{p1}{vpath1(__a0,__a5)}
\fmfiv{d.sh=circle,d.f=1,d.size=2thick,label=$a_1$}{point length(p1)/4 of p1}
\fmfiv{d.sh=circle,d.f=1,d.size=2thick,label=$a_2$}{point 2length(p1)/4 of p1}
\fmfiv{d.sh=circle,d.f=1,d.size=2thick,label=$a_3$}{point 3length(p1)/4 of p1}
\fmffreeze\fmfposition
\fmfipair{a[]}
\fmfiequ{a1}{point length(p1)/4 of p1}
\fmfiequ{a2}{point 2length(p1)/4 of p1}
\fmfiequ{a3}{point 3length(p1)/4 of p1}
\fmfi{plain}{vloc(__a0) -- a1}
\fmfi{plain}{a1 -- a2}
\fmfi{plain}{a2 -- vloc(__a5)}
\fmffreeze
\fmf{phantom}{a0,aa,a5}
\fmffreeze
\fmfi{double}{vloc(__aa) -- a2}
\end{fmfgraph*}
\end{fmffile} \\
${\displaystyle [2\pi i\,I(a_0;a_1;a_2)\cdot1]\otimes [1\cdot I(a_2;a_3;a_4)]\,,}$
\end{tabular}
\end{center}
Summing over the split polygons is equivalent to summing over all pairs of polygons contributing to the coproduct of the two multiple polylogarithms in the right-hand side of Eq.~\eqref{eq:Gonch_mono}. Thus, we obtain
\beq\bsp
\left(\cM_{a_{n+1}=a_i}\otimes\id\right)&\Delta(I(a_0;a_1,\ldots,a_n;a_{n+1}))\\
&\,
=(2\pi i\otimes 1)\,
\Delta\left(I(a_0;a_1,\ldots,a_{i-1};a_{i})\right)\,\Delta\left(I(a_i;a_{i+1},\ldots,a_{n};a_{n+1})\right)\\
&\,=
\Delta\left(\cM_{a_{n+1}=a_i}I(a_0;a_1,\ldots,a_n;a_{n+1})\right)\,,
\esp\eeq
which finishes the proof.


\end{document}